\documentclass[smallextended]{svjour3} 

\usepackage[utf8]{inputenc}
\usepackage[T1]{fontenc}

\usepackage{amsmath,amssymb,amsfonts}
\usepackage{array}
\usepackage{subcaption}
\usepackage{graphicx}
\usepackage{multirow}
\graphicspath{ {figs/} }
\usepackage{balance}
\usepackage[linesnumbered,ruled,vlined]{algorithm2e}
\let\oldnl\nl
\newcommand{\nonl}{\renewcommand{\nl}{\let\nl\oldnl}}
\usepackage{algorithmic}
\usepackage{esvect}
\usepackage{booktabs}
\usepackage{longtable}
\usepackage{tabularx}
\usepackage{anyfontsize}
\usepackage[all]{nowidow}
\usepackage[normalem]{ulem}
\usepackage{xcolor}
\usepackage{colortbl}
\usepackage{enumitem}
\usepackage{lineno}
\usepackage{natbib}

\newcolumntype{Y}{>{\centering\arraybackslash}X}

\definecolor{commentGray}{RGB}{120,120,120}
\renewcommand{\algorithmiccomment}[1]{\bgroup\color{commentGray}{//#1}\egroup}

\usepackage{framed}
\usepackage{tikz}
\definecolor{light-gray}{gray}{0.9}
\usepackage[framemethod=TikZ]{mdframed}
\mdfsetup{skipabove=5pt,skipbelow=3pt}
\usepackage{lipsum}
\mdfdefinestyle{RQFrame}{%
	linecolor=black,
	outerlinewidth=0.15pt,
	roundcorner=3pt,
	innertopmargin=2pt,
	innerbottommargin=2pt,
	innerrightmargin=4pt,
	innerleftmargin=4pt,
	backgroundcolor=light-gray}
\definecolor{javagreen}{rgb}{0.25,0.5,0.35} 

\newcounter{commentnumber}

\usepackage[hyphens]{url}
\usepackage[hidelinks]{hyperref}
\hypersetup{breaklinks=true}
\usepackage{amsthm}
\usepackage{newtxmath} 

\newcommand{\nsga}{NSGA-II\xspace} %
\newcommand{\nsgad}{NSGA-II-D\xspace} %
\newcommand{\ras}{random search\xspace}
\newcommand{\rs}{RS\xspace}
\newcommand{\mopso}{OMOPSO\xspace}

\newcommand{\metric}[0]{CID}
\newcommand{\metricexpanded}[0]{Coverage Inverted Distance}
\newcommand{\clarge}{$O_{Large}$\xspace} %
\newcommand{\cmid}{$O_{Medium}$\xspace} %
\newcommand{\csmall}{$O_{Small}$\xspace} %
\newcommand{\bclarge}{\boldmath${O_{Large}}$\xspace} %
\newcommand{\bcmid}{\boldmath${O_{Medium}}$\xspace} %
\newcommand{\bcsmall}{\boldmath${O_{Small}}$\xspace} %

\usepackage{soul}

\definecolor{codegreen}{rgb}{0,0.6,0}
\definecolor{codegray}{rgb}{0.5,0.5,0.5}
\definecolor{codepurple}{rgb}{0.58,0,0.82}
\definecolor{backcolour}{rgb}{0.95,0.95,0.92}

\usepackage[]{silence}        

\usepackage{nicefrac}

\usepackage{sansmath}

\usepackage{physics}

\usepackage[hyphens]{url}
\usepackage[hidelinks]{hyperref}
\hypersetup{breaklinks=true}
\usepackage{amsthm}
\usepackage{newtxmath} 

\usepackage[most, breakable]{tcolorbox}
\usepackage{tablefootnote} 

\begin{document}

\journalname{Empirical Software Engineering}

\author{Lev Sorokin \and Damir Safin \and Shiva Nejati}

\institute{Lev Sorokin \at {fortiss GmbH, Technical University of Munich}\\
\email{lev.sorokin@tum.de}\\
\and Damir Safin \at {fortiss GmbH, Technical University of Munich} \\
\email{safin@fortiss.org}\\
\and Shiva Nejati \at {University of Ottawa, Canada}\\
\email{snejati@uottawa.ca}}

\title{Can Search-Based Testing with Pareto Optimization Effectively Cover Failure-Revealing Test Inputs?}

\titlerunning{Limitations of Pareto-driven Search-based Testing}

\date{Received: date / Accepted: date
}

\maketitle
\thispagestyle{empty} 

\begin{abstract}

Search-based software testing (SBST) is a widely-adopted technique for testing complex systems with large input spaces, such as Deep Learning-enabled (DL-enabled) systems. Many SBST techniques focus on Pareto-based optimization 
where multiple objectives are optimized in parallel to reveal failures. However, it is important to ensure that identified failures are spread throughout the entire failure-inducing area of a search domain, and not clustered in a sub-region. This ensures that identified failures are semantically diverse and reveal a wide range of underlying causes. 
In this paper, we present a theoretical argument explaining why testing based on Pareto optimization is inadequate for covering failure-inducing areas within a search domain. We support our argument with empirical results obtained by applying two widely used types of Pareto-based optimization techniques,  namely \mbox{NSGA-II} (an evolutionary algorithm) and \mopso (a swarm-based algorithm), to two DL-enabled systems: an industrial Automated Valet Parking (AVP) system and a system for classifying handwritten digits. We measure the coverage of failure-revealing test inputs in the input space using a metric, that we refer to as the \emph{Coverage Inverted Distance} (\metric{}) quality indicator. Our results show that  \mbox{NSGA-II}  and \mopso are \emph{not} more effective than a na\"ive random search baseline in covering test inputs that reveal failures. We show that this comparison remains valid for failure-inducing regions of various sizes of these two case studies. Further, we show that  incorporating a diversity-focused fitness function as well as a repopulation operator in NSGA-II improves, on average, the coverage difference between NSGA-II and random search by 52.1\%. However, even after diversification, NSGA-II still does not outperform random testing in covering test inputs that reveal failures.   The replication package for this study is available in a GitHub repository~\citep{replication-package}.
\end{abstract}


\section{Introduction}
\label{sec:intro}

Search-based software testing (SBST) is an effective method for testing complex systems with large input spaces~\citep{Zeller17SBST}. SBST employs metaheuristics, such as evolutionary algorithms~\citep{Raja18NSGA2DT,BorgANJS21, humeniuk2022searchbased, Klück19Nsga2ADAS, Moghadam21Deeper, moghadam2023machine,  Riccio2020DeepJanus, sorokin2024guidingsearchsvm}, to reveal failure-revealing test cases. It formulates testing as an optimization problem, capturing system safety requirements through multiple objectives that are often optimized using Pareto-based algorithms.\\
Pareto-based SBST techniques are generally assessed by their effectiveness in identifying failures, focusing on the quantity and diversity of the failures detected. However, we still do not know if Pareto-based SBST is capable of achieving adequate coverage across the space of failure-revealing test inputs. Ideally, we are interested in a testing method that can achieve high coverage over the space of failure-revealing test inputs. That is, it can identify failure-revealing test inputs distributed across the entire failure-inducing area of a search domain, rather than clustered in one specific sub-area. \\ The identification of such diverse failing test inputs can support the detection of underlying failure causes and conditions leading to failures~\citep{Raja18NSGA2DT, JodatSpurious23}.
Existing research on software testing views the identification of diverse failures as an important testing objective~\citep{ AghababaeyanDiversity23,Feldt2016DiversityTC}. For example, 
Aghababaeyan et. al.~\citep{AghababaeyanDiversity23} argue that we are more likely to identify diverse failures of a system if we identify diverse test inputs.\\ \indent
Several Pareto-based SBST algorithms use heuristics to make the search more explorative with the original goal of escaping local optima. This approach is consistent with the broader focus within the evolutionary search community on balancing the trade-off between exploration and exploitation in search algorithms. By investigating various heuristics and hyperparameter tunings, researchers aim to adjust the behavior of algorithms across the spectrum of exploration and exploitation, which is crucial for enabling algorithms to avoid local optima~\citep{Crepinsek2013exploration}. Exploring the trade-off between exploration and exploitation has an important practical use case in software testing as well, since increasing exploration within often exploitative Pareto-based algorithms has been shown to lead to the discovery of more diverse failures. For example,  Clune et al.~\citep{MetaGAClune2005, CluneMutationCoverage08} propose making genetic operators in evolutionary algorithms more explorative by increasing mutation rates when the performance of the search shows marginal improvement. Multiple approaches combine Pareto-based optimization with novelty search~\citep{NoveltyMouret2011, Riccio2020DeepJanus, Zohdinasab23DeepAtash,Zohdinasab21DeepHyperion} and maximize the distance of candidate solutions to previously found test inputs to escape local optima in the fitness landscape. However, it has never been studied how well Pareto-based algorithms cover the failure-inducing regions within a search domain.

In this paper, we aim to study Pareto-based SBST techniques in terms of their capability to cover the failure-inducing test inputs within a search domain by addressing the following question:  
\vspace*{-.2cm}
\begin{tcolorbox}[colback=gray!10!white,colframe=black!75!black]
\textit{Can Pareto-based search-based testing achieve high coverage of failure-revealing tests?} 
\end{tcolorbox}


We first present a theoretical argument explaining why Pareto-based SBST algorithms cannot achieve high coverage of failure-revealing tests within the search space. Our argument builds on a definition of an SBST problem characterized by the following two  assumptions: 
\begin{itemize}[topsep=1em, parsep=1em,leftmargin=2.5em, itemsep=0.2em]\item[(\textbf{A1})]~The problem involves optimizing multiple quantitative fitness functions simultaneously using the principle of Pareto optimality. 
    \item[(\textbf{A2})]~The test oracle function of the SBST problem specifies subsets of the objective space as the failure regions such that at least one such subset has the same dimensionality as the objective space, i.e., the dimensionality is equal to the number of fitness functions.
\end{itemize}

For example, consider a SBST problem with two real-valued fitness functions $f_1$ and $f_2$, and a test oracle function $O$. Suppose  $O$ indicates a test input $v$ as a failure if the following constraint holds: $ (f_1(v) > 5) \vee (f_2(v) < 10)$. The test oracle $O$ satisfies assumption \textbf{A2} since it specifies a two-dimensional space as a failure region. Briefly, our argument indicates that any Pareto-based optimization algorithm for $f_1$ and $f_2$ ultimately identifies a Pareto front intersecting the two-dimensional failure region in only one dimension, which likely cannot ensure good coverage of the failure region defined by $O$.  Other than the assumptions \textbf{A1} and \textbf{A2}, our argument is not dependent on the number and definition of fitness functions, or  the heuristics used to approximate the optimal solution. We note that assumption \textbf{A2} is not constraining, as test oracles for many SBST formulations applied to cyber-physical systems and Deep Learning (DL) systems conform to this assumption~\citep{Raja16NeuralNSGA2,Raja18NSGA2DT, BorgANJS21,HamidSBTApollo21,JahangirovaOracles21}, making our argument valid for these existing SBST formulations. 


In addition to the theoretical argument, we investigate our research question empirically. We consider three Pareto-based algorithms that have been previously used for software testing: 1) \textit{\nsga}~\citep{Deb02NSGA2}, which is a genetic algorithm and has been widely applied for testing complex systems~\citep{Raja16NeuralNSGA2,Klück19Nsga2ADAS}; 2) an extension of NSGA-II that implements the concepts of novelty search combined with a diversification operator~\citep{Riccio2020DeepJanus}; and 3) the swarm-based Pareto-optimization algorithm \mopso~\citep{OmopsoSierra2005}, which is an extension of the well-known Particle Swarm Optimization algorithm for multi-objective problems.  We apply these three algorithms to two case studies representing two SBST problems: an industrial ADS case study Automated Valet Parking (AVP) system~\citep{BoschAVP} that is integrated in the Prescan simulator~\citep{Prescan}, and a hand-written digit-recognition Deep-Learning (DL) system~\citep{MNIST-model} that uses the MNIST dataset~\citep{MNIST} as input. We compare these three algorithms with a na\"ive random testing baseline, which is considered a standard baseline in the search-based software engineering community. In particular, we investigate whether the three algorithms can outperform naïve random testing in terms of covering the failure-inducing test inputs within a search domain. These algorithms or their close variants have already been compared with a random testing baseline regarding the quantity and diversity of failures, where failure diversity is measured, e.g., by calculating the average Euclidean distance between inputs~\citep{HamidSBTApollo21} or by applying the Jaccard similarity metric~\citep{HumeniukAmbiegen23}. However, there is no prior comparison with respect to the coverage of failure-inducing test inputs.

Although there are metrics for measuring diversity, none has been used to evaluate search algorithms by measuring how well they cover failure-inducing regions in the test input space. This is because measuring such coverage requires an approximation of the failure-inducing regions, which is a more difficult problem than finding individual failures. To measure the coverage of the failure-inducing areas of the input space, we propose to use a metric that we refer to as \textit{\metricexpanded{}} (\metric{}). We use this metric only to understand how different Pareto-based algorithms compare in terms of coverage. Otherwise, as we detail below, computing this metric may not be feasible in general. To compute \metric{},  we create a reference set that approximates 
failure-inducing regions within a search domain. Given such a reference set and given a solution set computed by a testing method, \metric{} measures how well the solution set covers the reference set by computing the average distance from each point in the reference set to the closest point in the solution set. The smaller the \metric{} value, the better the solution set in covering the space of all failure-revealing test inputs. To build the reference set, we generate many inputs evenly across the search space~\citep{Nabhan19coverage} such that these input points approximate the search space as a uniform grid. Among all the generated inputs, those that fail represent the reference set approximating the failure-inducing areas of a search domain.  While computing a reference set is feasible for our two case studies, it is likely infeasible for larger and more complex systems, as it requires running the system under test across numerous points in the search space. Hence, the \metric{} metric is not a metric that we propose to be used, in general,  for evaluating test generation algorithms for arbitrarily case studies.

The accuracy of the \metric{} metric depends on how well the reference set approximates the actual failure-inducing regions. To mitigate potential threats related to the accuracy of \metric{} in our study, we mathematically explain the relationship between \metric{}'s accuracy and the resolution of reference sets and specify conditions under which \metric{}'s error approaches zero. In addition, we present the useful properties of \metric{}, highlighting its robustness to the reference set, regardless of the method used to generate points in the input space for computing the reference set. Moreover, \metric{} is independent of the size of the test set generated by a testing approach, enabling the comparability of \metric{} values across test sets of different sizes.



\emph{\textbf{Results: }} For each of our two case studies, we define three alternative test oracle functions, each specifying a different failure region. These functions are defined to represent increasingly stricter failure criteria, allowing us to obtain experimental results that reflect varying difficulty levels in covering these regions in our case studies. Our results show that the three different Pareto-based testing techniques fare no better than na\"ive random testing in covering the input space regions that include failure-revealing test inputs:

\begin{itemize}
    \item  The results show that na\"ive random testing covers failure-inducing regions more effectively than \nsga as defined by all the alternative test oracles in the MNIST case study. For the AVP case study, na\"ive random testing surpasses \nsga\ for one test oracle, and for the other two test oracles specifying smaller and more restricted failure regions, both methods perform equally. These results indicate that \nsga cannot outperform  random search in terms of coverage of failure regions even when we vary the size of  failure-inducing regions.

\item Incorporating a diversity-focused fitness function and a re-population operator in \nsga improves, on average, the coverage difference between \nsga and random search by 52.1\%. However, \nsga, even after being extended by these diversification mechanisms, still cannot outperform random search in covering the failure-inducing regions of our two case studies and for the three different test oracle functions of these case studies.


\item The swarm-based search method, \mopso,  improves the coverage of failing test inputs compared to \nsga but, similar to \nsga, \mopso still does not perform better  than random search.
\end{itemize}

The outline of the paper is as follows: In Section~II, we present the preliminaries. In Section~III, we present related work, followed by our argument in Section~IV regarding why Pareto-based algorithms cannot achieve high coverage of failure-revealing tests. In Section~V, we introduce a coverage metric to assess the coverage of failures in the input space. In Section~VI, we compare, using our novel metric, Pareto-based driven testing with random testing in terms of the coverage of failure-inducing test inputs. In Section~VII, we discuss the most important threats concerning the validity of our results. In Section~VIII, we reflect on the lessons learned and conclude in Section~IX. In the appendix we provide proofs related to our introduced metric.

\section{SBST Problem Definition and Assumptions}
\label{sec:preliminaries}

In this section, we introduce the terminologies and formal notations  used for our argumentation in Section~\ref{sec:metrics}. We start with the definition of a SBST problem.

\begin{definition} \label{def:sbt}
   We define a search-based software testing (SBST) problem as a tuple $P = (S, D, F, O, SO)$, where
    \begin{itemize}
        \item $S$ is the system under test.
        \item $D \subseteq \mathbb{R}^n$ is the search domain, where $n$ is the $dimension$ of the search space. An element $x = (x_1, \ldots, x_n) \in D$ is called a test input. 
   \item $F$ is the vector-valued fitness function defined as \newline$F: D \mapsto \mathbb{R}^m$, \hbox{$F(x) = (f_1(x),\ldots, f_m(x))$}, where $f_i$ is a scalar fitness function (or fitness function for short) which assigns to each test input a quantitative value, and $\mathbb{R}^m$ is the \textit{objective space}. A fitness function evaluates how \textit{fit} a test input is. This captures the assumption \textbf{A1} described in Section~\ref{sec:intro}.
    \item $O$ is the test oracle function, $O : \mathbb{R}^m \mapsto \lbrace 0,1 \rbrace $, which evaluates whether a test  passes or fails. A test that fails is called \emph{failure-revealing}. We assume that 1 indicates fail and 0 indicates pass. 

    \item The set of \emph{all} failure-revealing test inputs is called the domain of interest (DOI). Note that $DOI$ is a subset of the search space (i.e., $DOI \subseteq D$). 
    The image of the domain of interest $F(DOI)$ is called codomain of interest (COI). Note that $COI \subseteq \mathbb{R}^m$, and for every $y \in COI$, we have O(y) = 1. 
     \item The test oracle $O$ is defined such that  COI has the same dimensionality as $\mathbb{R}^m$. This captures the assumption \textbf{A2} described in Section~\ref{sec:intro}.
    \item SO is the search objective. A search objective refers to a specific goal or criterion that guides the search process in finding test inputs that satisfy certain properties of the SUT (e.g., failure of a system).
    \end{itemize}
\end{definition}



As defined above, our definition of SBST problems involves multiple fitness functions. Solving such SBST problems requires a Pareto-based optimization algorithm.

\vspace{5pt}

\begin{definition}[Pareto-based Optimization] \label{def:moo}
A Pareto-based optimization problem is defined as 
\begin{align}
    \min_{x \in X} F(x) = (f_1(x), \ldots, f_k(x))
\end{align} where $f_i$ is an objective function and $X \subseteq \mathbb{R}^n$ is called the feasible solution set, where $n$ is the dimensionality of a search domain. 
A solution $x$ with $F(x) = (v_1,\ldots, v_m)$ is said to \textit{dominate} another solution $x'$ with $F(x') =(v_1',\ldots, v_m')$, if and only if $x$ is in at least one fitness value better than $x'$ and not worse in the remaining fitness values, i.e., $\exists v_i. (v_i < v_i') \wedge \forall v_j.(v_j \leq v_j')$. 
A solution $x$ is called Pareto optimal if no solution exists that dominates $x$.
The set of all Pareto optimal solutions is called Pareto set (PS), while the image of the Pareto set $F(PS)$ constitutes the Pareto front (PF).
\end{definition} 

As discussed in Section~\ref{sec:intro}, our argument in Section~\ref{sec:metrics} builds on two assumptions, \textbf{A1} and \textbf{A2}. Assumption \textbf{A1} states that the problem involves optimizing multiple quantitative fitness functions simultaneously using the principle of Pareto optimality, and assumption \textbf{A2} states that the test oracle $O$ is defined such that COI has the same dimensionality as $\mathbb{R}^m$. Both assumptions are included in Definition~\ref{def:sbt}. Both of our case studies, AVP and MNIST, satisfy assumptions \textbf{A1} and \textbf{A2}. Below, we illustrate how our AVP case study can be captured using our definition of an SBST problem, i.e., Definition~\ref{def:sbt}. Our second case study, which concerns testing a digit classification system, is described in Section~\ref{sec:setup}.

\textbf{Example.} Our first case study system is an AVP system~\citep{BoschAVP}. An AVP system is a feature added to a car (ego vehicle), which parks the ego vehicle without human intervention in a predefined parking spot by following a precalculated trajectory.
Our objective is to evaluate AVP system to verify if it meets the following safety requirement: 

$R$ =\emph{``When the ego vehicle's velocity exceeds threshold $th_1$, AVP must ensure a minimum distance of $th_2$ from both static and dynamic objects, including other vehicles and pedestrians.''} 

The values for parameters $th_1$ and $th_2$ are determined by the system's specific context and set by a domain expert. We execute AVP in a simulation environment, exposing the system to various situations known as 
\textit{scenarios}~\citep{Ulbrich15scenario}.
In particular, we test AVP using a scenario where a pedestrian crosses the ego vehicle's trajectory from the right of the ego vehicle's lane. We parameterize this scenario so that we can generate different instances of this scenario to stress the AVP system. We define the SBST elements from Definition~\ref{def:sbt} for AVP as follows:

\begin{itemize}
\item The search domain $D$ of AVP is defined by three search variables: $x_1$, the velocity of the ego vehicle; $x_2$, the velocity of the pedestrian; and $x_3$, the time when the pedestrian starts moving. Each search variable $x_i$ has a designated range $D_i$. Thus, the search space is defined as follows: $D = D_1 \times D_2 \times D_3$, where $D_1 = [0.3 m/s, 3 m/s]$, $D_2 = [0.5 m/s,2 m/s]$ and $D_3 = [0s,5s]$.
\item We define two fitness functions:
$f_1$ is the adapted distance between the ego vehicle and the pedestrian, and $f_2$ is the velocity of the ego vehicle at the time of the minimal distance between the ego vehicle and the pedestrian. 
An adapted distance function combines the longitudinal and latitudinal distances between the ego vehicle and a pedestrian into a single number, assigning more optimal values to test cases where the pedestrian is located in front of the car.
To compute a fitness value for a test input, AVP must be run in a simulation environment. The simulator provides outputs, including the position and velocity traces of both the ego vehicle and the pedestrian, from which the fitness values are calculated.

\item The test oracle function is defined as $O(x) = f_1(x) < th_1 \wedge f_2(x) < th_2$ where $x$ is a test input, and the parameters $th_1$ and $th_2$ are determined based on threshold values in the AVP safety requirement $R$ described earlier. The test oracle function evaluates whether, or not,  a test input violates the requirement $R$.



\item DOI is the set of all test inputs that lead to the violation of the safety requirement. That is, DOI is the set of all failure-revealing test inputs. COI, on the other hand,  is the set of the fitness values of the test inputs in DOI. We note that, in general, we cannot compute DOI for a given search-based testing problem. As we discuss in Section~\ref{sec:metrics_qi}, we propose a way to approximate DOI for our case studies. We refer to the approximation of DOI as a \emph{reference set}. 
\item The search objective $SO$ is to identify as good as possible the DOI, i.e., to identify diverse test inputs leading to the violation of the safety requirement. When we use a Pareto-based optimization algorithm in Definition~\ref{def:moo}, for testing, we often define the search objective as finding optimal fitness values. However, as motivated in Section~\ref{sec:intro}, our objective in this paper and for our case studies is to effectively cover DOI, i.e., the set of failure-revealing test inputs. 

\end{itemize}
\section{Related Work}
\label{sec:related-work}

In this section, we outline two streams of related work. In the first part, we discuss SBST approaches for generating diverse test data, highlighting both Pareto-based and non-Pareto-based methods. In the second part, we discuss metrics used to assess the diversity and coverage of SBST testing techniques.

\subsection{Diversified SBST Approaches}

Table~\ref{tab:related-overview-testing} provides an overview of SBST techniques aimed at generating diverse tests and failures. Below, we briefly discuss these approaches and contrast them with our work.


DeepJanus \citep{Riccio2020DeepJanus} is an \nsga-based search approach designed to identify the boundaries of the failure-revealing regions in DL-based systems. In particular, it extends a Pareto-based genetic algorithm with concepts from novelty search~\citep{Lehman2011NoveltySearch,NoveltyMouret2011}. While Pareto-driven testing promises identifying interesting tests by optimizing a single or multiple fitness functions, novelty search requires no definition of a fitness function. In novelty search the identification of promising solutions is guided solely by the \textit{novelty} a solution has. Novelty is quantified by using a distance measure applied between the candidate and solutions which have been identified in previous iterations of the algorithm (called archive). In addition, DeepJanus implements a repopulation operator, which replaces at every popoulation a portion of dominated individuals by newly generated candidates. The repopulation operator should diversify the search and support to escape local optima. 
In our study, in order to diversify \nsga, we have incorporated novelty search into the genetic algorithm of \nsga by using the diversity-aware fitness function of DeepJanus, which measures the Euclidean distance between a test input and the archive of previously found tests. In addition, we have used the repopulation operator from DeepJanus to increase the diversity of the solution set in our diversified \nsga.

Our results show that although incorporating a diversity-aware fitness function and the repopulation operator improves the average coverage difference between \nsga and random search, \nsga -- even when augmented with these diversification measures -- still cannot outperform random search in covering failure-revealing test inputs.

\begin{table}[t]
    \captionsetup{width=\textwidth}

    \centering 
    \caption{Classification of the related SBST approaches based on their underlying algorithm, the metric they use to measure the diversity of the generated tests, and whether or not they rely on Pareto-based optimization. FM stands for feature map; GA stands for genetic algorithm; and PSO stands for particle swarm optimization.}~\label{tab:related-overview-testing}
    \begin{tabular}{p{2.7cm}|p{3.5cm} | p{2cm} | p{2cm}}
         Reference & Algorithm &  Metric & Pareto-based (Y/N) \\ \hline
         DeepJanus~\citep{Riccio2020DeepJanus} & GA and Novelty Search  & Distance-based diversity & Y \\
         DeepAtash~\citep{Zohdinasab23DeepAtash} & GA and Novelty Search  & FM and Distance Archive  & Y \\
         DeepHyperion~\citep{Zohdinasab21DeepHyperion} & Illumination Search   & FM Diversity & N \\
                     \cite{LehmanNovelty11} & Novelty Search & Average Distance  & N \\
        \cite{Nabhan19coverage} & Evolution Strategy & Closest Point  & Y \\
        \cite{Moghadam21Deeper} & PSO  & Distance  & Y\tablefootnote{Extension to Multi-objective Optimization Problems is Pareto-based.} \\
        \hline
    \end{tabular} 
\end{table}

The approaches by Birchler et al. and Lu et al.~\citep{BirchlerPrio23, ChengjiePrio21} also apply a diversified fitness function to identify more diverse failures; however, these studies are concerned with test case prioritization and not test case generation, which is our objective.


DeepHyperion \citep{Zohdinasab21DeepHyperion} is a search-based testing approach adapting illumination search~\citep{Mouret15MAPElites} for testing deep-learning systems. Illumination search is an optimization approach that utilizes human-interpretable feature maps to identify failure-revealing test inputs. The feature space represents important and human-interpretable characteristics of test inputs. While test inputs are not directly generated in the feature space, they can be mapped to the feature space and positioned in the feature map. Based on already evaluated test inputs in the feature map, new test inputs are selected for subsequent search iterations. 
Since our paper focuses on studying the coverage of Pareto-based optimization algorithms and DeepHyperion is not a Pareto-based algorithm, we do not include its underlying algorithm in our study.
Further, the proposed metrics by Zhodinasab et al.~\citep{Zohdinasab21DeepHyperion} do not assess the coverage of failure-inducing regions. They propose two metrics: one for assessing the diversity of failures within the feature map and the other evaluates the diversity of the covered cells of the feature map. 
Their metrics are different from ours for two main reasons:
i)~First, they are defined over the feature space rather than the input space.  Therefore, metrics over the feature space can not directly indicate how much of the input space is explored/covered. ii)~Second, the second metric by Zhodinasab et al. does not approximate the set of actual failures as we do through our reference set. Hence, their metric cannot assess how well the identified failures represent the actual failures.






DeepAtash~\citep{Zohdinasab23DeepAtash} is a Pareto-based testing approach for deep learning systems. Like DeepHyperion, DeepAtash utilizes feature maps, enabling targeted searches in specific cells within the feature map to identify diverse solutions.  We do not include DeepAtash in our study because it is not designed to increase the diversity of identified failures; instead, it aims to target the identification of specific failures.

Moghadam et al.~\citep{Moghadam21Deeper} have 
presented multiple bio-inspired techniques for testing a DNN-based lanekeeping system. Their approaches are based on genetic algorithms GA and \nsga, evolution strategy and particle swarm optimization (PSO) and have been compared with the  Frenetic,
GABExploit and GABExplore, Swat techniques.
They particularly assess the diversity of identified failures by computing the average of the maximum Levenshtein distance between road segments that trigger failures. Their results indicate that their presented approaches have comparable performance in terms of producing diverse test inputs. The study does not evaluate the coverage of the failures in the DNN's input space, but only the diversity of the identified failures. Instead, in our work, we focus on the coverage of identified failures. Among the algorithms studied by Moghadam et al., we evaluate the performance of \mopso, which is an extension of the particle swarm algorithm to multi-objective optimization problems and therefore a Pareto-based algorithm.



The closest work to our work is the study from Nabhan et al. \citep{Nabhan19coverage}. Similar to our work, they evaluate the coverage of the failure-inducing regions of the input space  by approximating the failure-inducing test input space - the DOI - using grid sampling. To assess how well a solution set of an algorithm covers the failures of a systems, they calculate the portion of failure-revealing grid points that have a witness in the solution set. A grid point is said to have a witness in the solution set if the distance to the closest test input within the set does not exceed a fixed threshold.
However, Nabhan et al.'s paper does not study Pareto-based testing techniques in terms of the coverage of the test input space, which is the primary focus of our work.

\subsection{Diversity and Coverage Metrics for SBST}
Several metrics are proposed to assess Pareto-based testing algorithms with respect to the objective space, which is the space of fitness/objective values~\citep{LiEvaluateSBST22, LiQuality2019}. In contrast, since in our paper we focus on the input space, we have included in Table~\ref{tab:related-overview-cases} an overview of the approaches proposed to assess the capabilities of SBST techniques in terms of diversity and coverage when the focus is either on the input space or the feature space, i.e., a representation of the input data.

Below, we discuss these techniques and contrast the metric they use to measure diversity or coverage with our Coverage Inverted Distance (CID) metric.

\begin{table}[t]
    \captionsetup{width=\textwidth}

    \centering
    \caption{Classification of existing studies assessing test case diversity and coverage of SBST algorithms based on the following criteria: their underlying search algorithms, the metrics they use to measure diversity or coverage, whether the metric measures diversity, and whether the metric measures coverage.    
    RS stands for Randomized Search, GA stands for Genetic Algorithm, and NMCTS stands for Nested Monte Carlo Tree Search.}\label{tab:related-overview-cases}
    \begin{tabular}{p{2.5cm}|p{2.5cm} | p{2.7cm} | p{1cm}  | p{1cm} }
         Reference & Search Algorithm(s) & Metric & Diversity (Y/N) & Coverage (Y/N)\\ \hline
                  \cite{HamidSBTApollo21} &  RS and GA & Euclidean Distance  & Y & N\\
                           \cite{humeniuk2022searchbased, HumeniukAmbiegen23,Humeniuk24Reinforcement} & NSGA-II and RS & Jaccard Similarity & Y & N\\
         \cite{feldt2017featurediversity}  &  RS, Hill-Climbing and NMCTS& Feature Map Coverage & Y & Y\tablefootnote{The work considers FM coverage, while we consider failure coverage in the input space.}\\
       \cite{Feldt2016DiversityTC}  &  N/A & Normalized Compression Distance & Y & N\\
          \cite{Hungar2020exploration}  &  N/A & Coverage Criterion & N & Y\\
                   \cite{ramakrishna2022riskaware}  &  N/A & Cluster-based & Y & N\\
         \cite{Biagiola2019DiversityWeb} & RS & Numeric/String Distance & Y & N\\
         \cite{AghababaeyanDiversity23} & (Metric Study)  & Geometric Diversity & Y & N\\
         \cite{Neelofar24KeyFeatures, Neelofar24Adequacy} & (Metric Study) & Instant Space Metric & Y & N\tablefootnote{It is claimed that coverage is assessed, but no reference set is used.}\\ 
         This paper & NSGA-II, NSGA-II-D and \mopso & CID & N & Y \\ \bottomrule
    \end{tabular} 
\end{table}



Feldt et al.~\citep{Feldt2016DiversityTC} have proposed a metric to evaluate the diversity of a test suite. The metric is based on the Normalized Compression Distance (NCD), which is a distance metric based on information theory and can be applied to any type of test data. However, this metric is not meant for assessing the test input space coverage (DOI), but rather the diversity of test data. While diversity correlates with coverage, in our approach, we focus on assessing the coverage of failure-inducing regions with the goal of understanding the capabilities of Pareto-based algorithms in covering these regions.

Humeniuk et al.~\citep{HumeniukAmbiegen23, Humeniuk24Reinforcement} have evaluated the diversity of generated tests on a lane-keeping system case study using Pareto-optimization-based search (NSGA-II) compared to random search (RS). Using their metric based on Jaccard similarity, they have shown that, contrary to several prior studies, RS indeed generates more diverse tests than NSGA-II. While in our work, we study coverage instead of diversity, the findings of Humeniuk et al. are in line with our argument and our empirical results, demonstrating the weakness of Pareto-based optimization in covering failure-inducing regions of the search space.

Aghababaeyan et al. \citep{AghababaeyanDiversity23} have argued that diverse test inputs most likely point to diverse faults and evaluated three different metrics 
in assessing the diversity of image data used for testing deep learning models. They identify the geometric diversity relation as the most expressive metric. Unlike our metric, the metrics proposed by Aghababaeyan et al. do not evaluate the coverage of failures~\citep{AghababaeyanDiversity23}.

Biagiola et al.~\citep{Biagiola2019DiversityWeb} devised an SBST algorithm to maximize diversity in test inputs for generating test cases for testing web applications. However, the approach targets code coverage (branch/statement coverage) rather than the coverage of failures, which is why the proposed coverage evaluation technique is not applicable to our study.





Hungar et al. \citep{Hungar2020exploration} have noted the importance of covering failure-revealing test input spaces for ADS. They propose a binary coverage criterion for the search space. The criterion provides an assessment as either covered or not covered, and therefore, cannot be used as a quantitative measure of DOI coverage.


Ramakrishna et al. \citep{ramakrishna2022riskaware} define a diversity metric that is based on clustering of ADS testing results. This metric is defined by the number of clusters and includes a $risk\ score$ for each test input or scenario. The approach requires the number of clusters to be provided as input to the clustering technique. However, the process for determining the number of clusters is unclear, as is the potential impact of this predetermined number of clusters on the diversity assessment results.
Marculescu et al. \citep{Marculescu16} focus on the exploration of the objective space and 
compare exploration-based evolutionary search algorithms such as novelty search and Illumination Search (IS) with Differential Evolution (DE), an optimization algorithm. Their results show that IS and DE explore larger and different regions of the objective space (COI in our context) compared to solely optimization-based algorithms. Nevertheless, this study does not evaluate the coverage of the DOI in particular when using Pareto-based search approaches, which is the focus of our work.


Feldt and Poulding~\citep{feldt2017featurediversity} have studied the feature space~\citep{Mouret15MAPElites} coverage when using randomized, optimization-based, and hybrid testing approaches such as Nested Monte Carlo Tree Search \citep{MCTSBrowne12}. Their findings show that randomized testing exhibits similar coverage results compared to pure optimization-driven testing. However, their notion of coverage differs from ours. Their metric is similar to that of Zhodinasab et al.~\citep{Zohdinasab21DeepHyperion}, which is focused on the feature space rather than the input space and does not approximate the set of actual failures as we do through our reference set.

Ebabi et al.\citep{HamidSBTApollo21} have investigated the diversity of generated test data when testing an ADS with a genetic algorithm compared to random search. By evaluating the average Euclidean distance of identified failing tests, their findings show that genetic search outperforms randomized search in producing diverse test inputs. However, they do not assess the coverage abilities of the testing techniques compared to our study. In particular, their results deviate from the study by Humeniuk et al.\citep{HumeniukAmbiegen23}, which has shown that RS outperforms \nsga in terms of the diversity of test inputs.

Neelofar et al.~\citep{Neelofar24Adequacy, Neelofar24KeyFeatures} have proposed a set of adequacy metrics to assess the diversity and coverage of test suites for testing DL-enabled system. The metrics project test data from a multidimensional feature space into a two-dimensional space called instance space and evaluate coverage and diversity using information-theoretic metrics. However, similar to the coverage metrics proposed by Feldt and Poulding~\citep{feldt2017featurediversity} and Zhodinasab et al.~\citep{Zohdinasab21DeepHyperion}, in these papers, the coverage is not assessed by contrasting the identified failures with the actual failures since these approaches do not approximate the set of actual failures as we do through our reference set. Without knowing the space of actual failures, we cannot provide any information about the coverage of the failure-inducing space. These existing coverage metrics only compare the space of identified failures with the entire feature or instant spaces. This information, while useful, does not indicate what portion of the actual failures are found by a testing algorithm. Our study, for the first time, attempts to address this question for Pareto-based optimization techniques.


\section{Theoretical Argumentation}
\label{sec:theoretical}

We claim that Pareto-optimal solutions computed by Pareto-based approaches do not necessarily lead to an adequate coverage of the DOI defined in Definition~\ref{def:sbt}, i.e., the set of failure-revealing test inputs. Our argumentation builds on the definitions of the SBST problem (Definition~\ref{def:sbt}) and Pareto-based optimization (Definition~\ref{def:moo}).


\begin{figure*}
    \centering
    \begin{subfigure}[t]{0.48\textwidth}
      \centering
\includegraphics[width=0.5\linewidth]{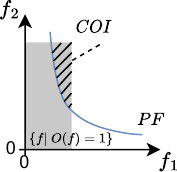}    \caption{}
\label{fig:example_argumentation_os} 
    \end{subfigure}
    \hfill
    \begin{subfigure}[t]{0.48\textwidth}
   \centering\includegraphics[width=0.5\linewidth]{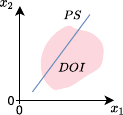}
    \caption{}
    \label{fig:example_argumentation_ds}
    \end{subfigure}
    \caption{Illustrating the limitation of Pareto-based optimization algorithms in covering failures: Figure a) illustrates how a Pareto front (PF) and a codomain of interest (COI) are located in the objective space. Figure b) shows how a Pareto set (PS) and a domain of interest (DOI) are located in the search space. The dimensionalities of PF and PS are, respectively, lower than the dimensionalities of COI and DOI. Hence, PF cannot effectively cover COI, and PS cannot effectively cover DOI. }\label{fig:example_argumentation}
    
\end{figure*}

We use Figures~\ref{fig:example_argumentation_os} and~\ref{fig:example_argumentation_ds} to illustrate our argument. These figures, respectively, show 
the objective and search spaces for a Pareto-based optimization problem with two fitness functions, i.e., $m=2$, and two input variables, i.e., $n=2$. 
The grey area in Figure~\ref{fig:example_argumentation_os} represents the test oracle function $O$ which identifies the subset of the objective space encompassing failures.  The portion of the grey area dominated by the true Pareto-Front (PF)  and delineated with dashed lines in Figure~\ref{fig:example_argumentation_os} represents  the set of reachable failures which is called the codomain of interest (COI) in Definition~\ref{def:sbt}. Due to assumption \textbf{A2} in Definition~\ref{def:sbt}, we assume that COI has the same dimensionality as the objective space. That is, the number of dimensions of both COI and the objective space is $m$ which is two in our example.  The PS line and the DOI area in Figure~\ref{fig:example_argumentation_ds}, respectively, show the mappings of PF and COI in the search space, i.e., input space.
Assuming that there is no redundancy in search variables and fitness functions, which is achievable through global sensitivity analysis~\citep{MatinnejadNBB14}, the DOI has the same number of dimensions as the search space. That is, the number of dimensions of both DOI and the input space is $n$ which is two in our example. However, PF, computed within $\mathbb{R}^m$, has at most $m-1$ dimensions. Consequently, PF with at most $m-1$ dimensions cannot effectively cover a COI with $m$ dimensions. Similarly, PS, which maps PF to DOI, has at most $n-1$ dimensions. Therefore, PS with at most $n-1$ dimensions cannot effectively cover a DOI with $n$ dimensions.



\emph{In summary,} using a Pareto-based optimization algorithm results in computing PS in the search space and PF in the objective space. However, PS cannot achieve good coverage of DOI, just as PF falls short in covering COI, as evidenced by our argument above and the illustrations in Figures~\ref{fig:example_argumentation_os} and~\ref{fig:example_argumentation_ds}. Therefore, a Pareto-based optimization algorithm is not suitable for achieving adequate DOI coverage.

\section{Metric for DOI Coverage Assessment}
\label{sec:metrics}
To empirically assess the coverage of the actual failing test inputs of an SUT, we introduce the \metric{} metric in this section. \metric{} allows us to quantitatively evaluate the coverage of the DOI by failure-revealing test inputs. In Section~\ref{sec:metrics_qi}, we introduce \metric{}. In Section~\ref{sec:metrics_refset}, we explore methods for approximating the set of failures, which serves as the reference set for \metric{}, and discuss the accuracy and robustness of reference sets.

\subsection{\mbox{Metric Definition}}
\label{sec:metrics_qi}

In this section, we present \textit{\metricexpanded{}} (\metric{}), a metric specifically designed to assess how well the failures identified by a testing technique cover the actual set of failure-revealing test inputs for a system. That is, \metric{} evaluates how well a given set of failures identified by a testing technique covers the DOI. Let $Z$ be a finite set approximating the DOI for a given SBST problem, and let $A$ be a finite set of solutions generated by a testing method for that SBST problem. In the remainder, we call $Z$ a reference set and $A$ a test set. Our metric \metric{}, defined below, assesses how well the test set A approximates, i.e. covers, the reference set $Z$:

\begin{align}
     \metric{}(A,Z) = \frac{1}{\left \lvert Z \right \rvert} \left( {\sum_{z \in Z} {d_z}^q} \right)^{\nicefrac{1}{q}}
     \label{eq:metric}
\end{align}

, where $d_z = \left( \sum_{j=1}^{n} {|z_j - x_j|}^p \right)^{\nicefrac{1}{p}}$ represents the distance imposed by the p-norm from a reference point $z = (z_1, \ldots, z_n) \in Z$, to the nearest point $x = (x_1, \ldots, x_n)$ from the test set $A$ in the search space, while $q$ represents the norm for computing the mean of the distance (i.e., $q=1$ for the generalized mean, or $q=2$ for the power mean). If $q=1$ is chosen, the metric results in the average distance from a reference point to the closest point in $A$. If at the same time $p=2$ is chosen, then the defined distance between the points is the Euclidean distance. If no specific characteristics of the system under test are known, the Euclidean distance is the most practical 
option~\citep{DistanceFunctions}. The lower the \metric{} value, the better the test set represents the reference set. The \metric{} value reaches zero if the test set achieves a complete representation of the reference set. 

 In our evaluation in Section~\ref{sec:application}, we use the Euclidean distance to compute the distance between the points in the reference set ($Z$) and  the test points in the solution set ($A$), i.e., $p=2$. Further, we use the general mean to compute CID values, i.e, $q=1$.

\begin{figure*}
    \centering
    \begin{subfigure}[b]{0.25\textwidth}
        \label{fig:cigd_example_1}
        \centering
        \includegraphics[width=\textwidth]{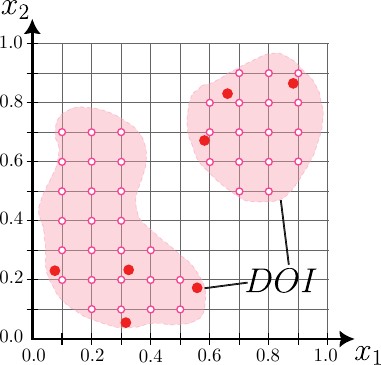}
        \caption{}
    \end{subfigure}
    \hfill
    \begin{subfigure}[b]{0.25\textwidth}
        \label{fig:cigd_example_2}
        \centering
        \includegraphics[width=\textwidth]{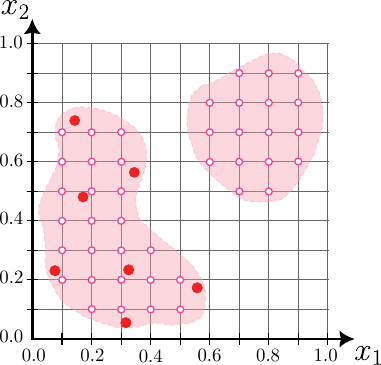}
        \caption{}
    \end{subfigure}
    \hfill
    \begin{subfigure}[b]{0.25\textwidth}
        \label{fig:cigd_example_3}
        \centering
        \includegraphics[width=\textwidth]{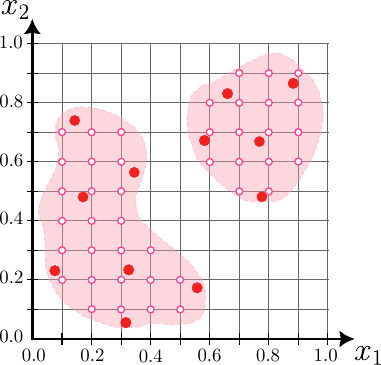}
        \caption{}
    \end{subfigure}
    \caption{Illustrative examples showing the computation of \metric{}: DOI is represented as two separate regions, highlighted in pink in the three examples. The test set, $A$, is presented as solid red points in all three examples. The reference set, $Z$, is represented by the non-filled pink circles. The example in (a): $A$ barely covers DOI; the example in (b): $A$ covers one region of DOI well; and the example in (c): $A$ covers both regions of DOI well. }
    \label{fig:cigd_metric_example}
    \vspace*{-.2cm}
\end{figure*}

While $A$ is computed by a given testing method, the reference set $Z$ should be computed independently from the testing method, and should represent the DOI as accurately as possible. 
In our study, we approximate the reference set $Z$ using a \textit{grid sampling} approach. Specifically, we discretize the search domain by segmenting each search variable into a predetermined number of equal intervals. We test the system under test, e.g., ADS, for each point in the discretized search domain to determine whether it is passing or failing. The reference set $Z$ is then defined as the set of failing test cases within the search space.\\
\indent We discuss \metric{} and its adequacy using illustrative examples. Figure~\ref{fig:cigd_metric_example} illustrates three examples, in which the DOI is represented as two separate regions, highlighted in pink, and the test set $A$ and reference set $Z$ are illustrated as solid red, and empty pink circles respectively. In Figure~\ref{fig:cigd_metric_example}(a), $A$ includes a handful of points in each of the regions and provides low coverage of both. In Figure~\ref{fig:cigd_metric_example}(b), $A$ contains several points exclusively within one region, providing a good coverage in one region, while not witnessing the other region. In Figure~\ref{fig:cigd_metric_example}(c), $A$ includes multiple points well-distributed in both regions, leading to a good coverage of the entire DOI. The reference set is the same for all three examples, and is generated by performing grid sampling with a step size of $0.1$ for each axis interval. For the first and second examples, \metric{} values are equal to 0.163 and 0.237 respectively, which indicates poor coverage of DOI by the corresponding test sets. For the third example, \metric{} is equal to 0.084, reflecting a good DOI coverage.

\subsection{\mbox{Reference Set Generation and Properties of CID}}
\label{sec:metrics_refset}

To apply CID, we need to generate a reference set to approximate the actual set of failures exhibited by the SUT. We can use various sampling techniques to create test inputs that are uniformly scattered across the search space in order to generate this reference set. Note that when we sample points to generate the reference set, we generate values within the search space. Sampling in this context means generating a set of points in the search space according to a sampling strategy. One such strategy is grid sampling, discussed in Section~\ref{sec:metrics_qi} where the selected test inputs are the nodes of a regular grid within the search space. Other alternative sampling strategies are \emph{furthest point sampling}~\citep{FPS} and \emph{Poisson disc sampling}~\citep{BridsonPDS07}. FPS is a sampling strategy that iteratively generates samples in such a way that the minimal distance from already sampled points to a new sample is maximized. Poisson disc sampling allows generating points in a given space while maintaining a user-defined minimal distance between sampled points.

We call a reference set generated by a uniform sampling strategy, such as the three strategies above,  a \emph{uniform reference set}. Consider a DOI containing several disconnected regions.   We say a uniform reference set is \emph{optimal} for a DOI if all the separate regions of the DOI are covered by the reference set.  We can show, that for an optimal reference set, the error of \metric{} reduces linearly as the maximal distance between adjacent points in the reference set decreases.
As we increase the number of points sampled with a uniform sampling approach, the \metric{}'s error tends to zero.
The sketched proofs of the above statements are available in the appendix. Further, the \metric{} metric has the following two important properties: \begin{itemize}
    \item[1.] \metric{} is not impacted by the size of the test set. This is because in the definition of \metric{} in Equation~\ref{eq:metric}, $d_z$ is computed as the distance from each reference point to the closest test point, and averaged over the reference set. This property enables us to compare test sets with different sizes;
    \item[2.]\metric{} values are robust when we use different reference sets as long as the reference sets are optimal and have equal sizes.
\end{itemize} 


Note that generating a reference set for an SBST problem 
requires executing the system under test for each sampled point, which can be computationally expensive, particularly for complex systems where each execution may take some minutes. In this paper, we use \metric{} to show the limitation of Pareto-based search-based testing for our DL-enabled case study systems with manageable search domain size and execution time.The applications of \metric{} are further discussed in Section~\ref{sec:lessons}.

\section{Empirical Study}
\label{sec:application}
While Section~\ref{sec:theoretical} presents an argument addressing our research question introduced in Section~\ref{sec:intro}, this section aims to answer our research question through empirical evidence. We assess the effectiveness of Pareto-based testing algorithms by comparing their ability to achieve high coverage of failure-revealing tests with that of baseline random testing. To ensure the validity and diversity of our empirical results, our experiments include alternative Pareto-based testing algorithms, case study systems, and varying definitions of test oracles for these systems.

In particular, we selected three optimization testing algorithms or their variants, which are Pareto-based and have been previously applied to case studies similar to ours, as identified in our literature survey in Table~\ref{tab:related-overview-testing}. Specifically, we selected the genetic algorithm \nsga, DeepJanus, a hybrid genetic algorithm that uses concepts from Novelty Search~\citep{NoveltyMouret2011} and evolutionary search, and \mopso~\citep{OmopsoSierra2005}, a swarm optimization algorithm. Other algorithms have been not considered in our study as they are non Pareto-based and hence outside of our scope. We present our empirical results using the following two sub-RQs:

\textbf{RQ\textsubscript{1}: How does Pareto-driven search-based testing compare to baseline random testing in terms of covering failure-revealing tests?}

To answer this research question, we use our proposed \metric{} metric to compare the test results obtained by the well-known Pareto-based algorithms NSGA-II, \mopso and a na\"ive testing algorithm random search (RS) in terms of DOI coverage. NSGA-II has been extensively used in test automation for ADS and DL-enabled systems~\citep{Raja16NeuralNSGA2,Raja18NSGA2DT,Klück19Nsga2ADAS,Riccio2020DeepJanus}. Similarly, \mopso is a Pareto-based swarm optimization technique that extends the well-known Particle Swarm Optimization (PSO) algorithm to solve multi-objective optimization problems. PSO is a swarm intelligence algorithm that leverages concepts from collective social behavior to identify solutions with optimal fitness values. PSO has been widely applied in various domains, ranging from image segmentation in image processing to solving wireless communication optimization problems~\citep{Shami22PSOSurvey}, and it has also been used for testing ADS~\citep{Moghadam21Deeper}.
We apply NSGA-II, \mopso and random testing to two case studies:  an industrial Automated Valet Parking system introduced in Section~\ref{sec:preliminaries}, and an open-source, DL-based, handwritten digits classification system that is widely-used for evaluating testing approaches in the literature~\citep{Riccio2020DeepJanus, Zohdinasab23DeepAtash, Zohdinasab21DeepHyperion}. We refer to the first case study as AVP, and to the second case study as MNIST since it uses digits from the MNIST dataset as test inputs. Both case studies have predefined test oracle functions for identifying failures. We extend the existing oracle functions by developing alternative definitions of test oracle functions, with some being stricter than
others. Stricter functions result in smaller DOI regions. We evaluate the DOI coverage performance of \nsga, \mopso and RS for the alternative test oracle functions to determine if and how the size of failure regions affects the coverage of failure tests by \nsga and RS.


\textbf{RQ\textsubscript{2}:How does state-of-the-art, diversity-focused, Pareto-driven search-based testing compare with random testing in terms of covering failure-revealing tests?}

Several Pareto-based SBST algorithms aim to achieve diversity and coverage of different failures by incorporating a diversity fitness function. In our study, we consider a diversified search algorithm similar to the proposed technique DeepJanus~\citep{Riccio2020DeepJanus}, which combines a genetic algorithm with concepts from novelty search. Specifically, we extend \nsga using a repopulation operator as well as a diversity-focused fitness function derived from DeepJanus, which aims to maximize the Euclidean distance of a test input to previously found solutions. Similar to RQ1, we use the \metric{} metric to compare the diversity-focused \nsga against RS with respect to DOI coverage.







\subsection{Experimental Setup}
\label{sec:setup}
In this section, we describe our experimental setup including the details of our case study systems, the implementation of \nsga, diversified \nsga, denoted by \nsgad, \mopso, and RS, and the metrics used in our empirical analysis. An overview of the configuration of the search algorithms is given in Table~\ref{tab:config_algos}. We note that we selected the parameters in this table after conducting preliminary experiments aimed at identifying optimal parameters for our three studied algorithms, \nsga, \nsgad, \mopso.

\begin{figure}[t]
    \centering
    \includegraphics[scale=0.6]{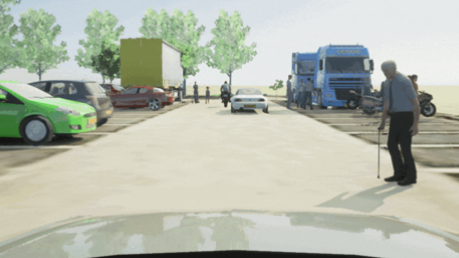}
    \caption{The test scenario for the AVP Case Study: A pedestrian is crossing the trajectory of a vehicle equipped with the automated valet parking system.}
    \label{fig:avp-cs}
\end{figure}

\begin{figure}[!tbp]
    \centering
    \begin{tabular}{c c}
         \includegraphics[width=0.1\textwidth]{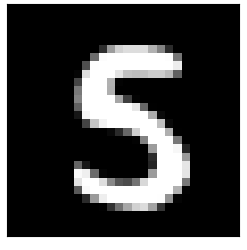} &     \includegraphics[width=0.105\textwidth]{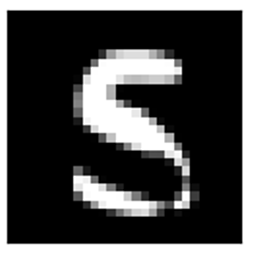} \\
      \includegraphics[width=0.1\textwidth]{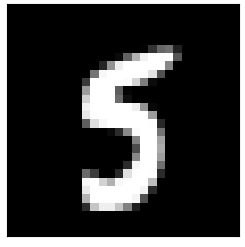} &
    \includegraphics[width=0.105\textwidth]{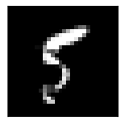} \\
    \includegraphics[width=0.1\textwidth]{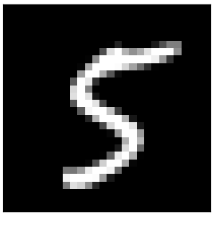} &
        \includegraphics[width=0.105\textwidth]{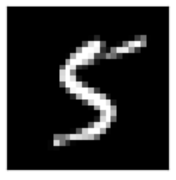}
    \end{tabular}
    
    \caption{Example test inputs for the MNIST case study. Left column: original digits, correctly classified as 5. Right column: corresponding label-preserving digits generated by \nsga and labeled as 8 by the classifier under test. Classification certainty differences between the expected label and the highest prediction by the classifier for the mutated digits from the top right to the bottom right are: -0.51, -0.79, -0.69.}

    \label{fig:mnist-cs}
\end{figure}

 \textbf{SUT MNIST.} 
 Our first case study focuses on a classification task involving the utilization of DL models. Specifically, we evaluate a DL model designed to classify handwritten digits and trained using the widely-used MNIST dataset~\citep{MNIST}. The SUT must accurately classify a digit within a 28x28 greyscale image, where the digit is depicted with white strokes on a black background. Our objective is to create new instances of MNIST digits that lead to a misclassification by the classifier (\autoref{fig:mnist-cs}). 

\textbf{SUT AVP.}
 The SUT of the second case study is the AVP system described in Section~\ref{sec:preliminaries}. AVP consists of a Lidar-based object detector, a static path planner for the generation of a driving path to a free parking lot, and a path follower. Further, it contains basic (longitudinal/latitudinal) vehicle dynamics. 

 \textbf{\nsga and RS for MNIST.} For the first case study, similar to existing work~\citep{Riccio2020DeepJanus}, we represent the input images as B\'ezier curves in the SVG format where the curve is characterized by several start points, end points, and control points. The mutation of digits throughout the search is performed by choosing a start point, an end point or a control point of the B\'ezier curve and applying a small tweak. The value of the tweak is chosen randomly from a predefined range. To select the range, we performed some preliminary experiments to identify a range size such that tweaking a digit less likely leads to the generation of an invalid digit. An invalid digit is one that cannot be recognized as a digit by human~\citep{Riccio23InputValidity}.

After mutating the Bezier curve, it is converted back to an image of size
28x28 and passed to the SUT for classification. For further technical details of how the transformations are enabled, we refer the reader to the paper by Riccio and Tonella~\citep{Riccio2020DeepJanus}. To generate the initial population, we select one seed image from MNIST dataset and apply the same mutation discussed above. In addition, we ensure that the distance of the mutated digits and the original seed image is less than a threshold proposed in existing work~\citep{Riccio2020DeepJanus,Zohdinasab21DeepHyperion}. This increases the likelihood of preserving the label of the initial seed for images in the initial population. As system under test, we consider the deep convolutional neural network provided by Keras\footnote{https://keras.io/} and used in existing studies~\citep{Riccio2020DeepJanus,Zohdinasab21DeepHyperion}.


  
We use two fitness functions to increase chances of generating images that are challenging for the digit classifier: $f_1$, adopted from previous studies \citep{Zohdinasab21DeepHyperion, Riccio2020DeepJanus}, represents the difference between the confidence in the expected label and the highest confidence among other labels.
We want to minimize $f_1$ since a lower $f_1$ value indicates a stronger tendency to missclassify a digit. Using $f_2$, we minimize the average brightness of generated digits so that they are more difficult for the digit classifier. For example, Figure~\ref{fig:mnist-cs} shows seed images, representing the digit 5, from the MNIST dataset on the left. On the right side, the figure displays corresponding mutations of these images with reduced average brightness, hence posing a challenge for digit classifiers. The goal of fitness $f_2$ is to increase the chances of generating images like those on the right side of Figure~\ref{fig:mnist-cs}. We calculate $f_2$ by dividing the sum of the brightness values of pixels, which have brightness values above zero, by the number of these pixels. For instance, the $f_2$ values for the digits in the first row in Figure~\ref{fig:mnist-cs} are 0.82 and 0.77, respectively, and in the second row 0.79 and 0.66.

\textbf{\mopso for MNIST and AVP.} For the configuration of the \mopso algorithm, we considered guidelines from existing works that have applied \mopso to different case studies~\citep{Shami22PSOSurvey, Shi1999InertiaWeight, Coello2004HandlingMO, OmopsoSierra2005}. In particular, for MNIST and AVP, we set the crowding distance archive size as the swarm size and used the same population size value as for the search with \nsga and \nsgad. Another component of the \mopso algorithm is the \textit{inertia weight} parameter, which controls how explorative or exploitative the particles are. A high inertia weight value results in a greater change in the position/test inputs of particles, potentially skipping interesting/failure-revealing regions, while a low value might hinder exploration and the detection of new failing tests. We compared the performance of the \mopso search using a static inertia weight with a time-dependent (dynamic) inertia weight~\citep{Shi1999InertiaWeight}, where the inertia weight is reduced with increasing iteration numbers. A dynamic inertia weight yielded better coverage results in preliminary experiments, which is why we chose to proceed with this configuration.
In addition, we investigated the impact on the search results when using different boundary handling techniques, i.e., absorbing and reflecting particles at the boundaries. In the first technique, particles are absorbed at the boundary, where their positions are set to the boundary value with velocity 0 whenever they reach a position outside the search space. In contrast, a reflecting boundary handling strategy reflects the particle back into the search space by inverting its velocity component whenever it reaches a position outside the search space. For our study, we selected the reflecting technique, as with the former approach, we observed that particles were stuck at the boundaries, exhibiting lower coverage scores.

\textbf{Test Oracles for MNIST.} We define three different test oracle functions for the MNIST case study, listed from least to most strict. These correspond to a decreasing size of the DOI region for this case study. Let $x$ be a test input. We define three test oracle functions for the MNIST case study as follows: 
\setlength{\abovedisplayskip}{12pt} 
\setlength{\belowdisplayskip}{12pt} 
\begin{align*}
    O_{Large} (x):  f_1(x) < -0.5 \\
    O_{Medium} (x):  f_1(x) < -0.7 \\
   O_{Snall} (x):  f_1(x) < -0.95
\end{align*}

Specifically, according to \clarge, a test input fails when the difference between the expected label and the highest prediction certainty generated by the classifier, i.e., the value of the fitness function $f_1$, is just less than $-0.5$. Functions \cmid and \csmall, on the other hand, require this difference to be less than $-0.7$ and $-0.95$, respectively. Hence, \cmid is stricter than \clarge, and \csmall is the strictest. 

We set the search budget for both \nsga and RS to 1,000 evaluations, as preliminary experiments have shown that for \nsga, \metric{} values stabilize already before. We use the population size of 20 and set the number of generations to 50.


 \textbf{\nsga and RS  for AVP.} In the second case study, we test the AVP system on a driving scenario where an occluded pedestrian is crossing the planned trajectory of the ego vehicle~(Figure~\ref{fig:avp-cs}). An occluded pedestrian refers to a pedestrian who is hidden from the ego vehicles's view because the pedestrian is either totally or partially behind another static object.
 The search space and fitness functions for  AVP are defined in Section~\ref{sec:preliminaries} and has been provided by our industrial partner. To adjust the velocity of the ego vehicle we just use a scalar value ranging from 0.1 to 1 which is internally used in the SUT to calculate the vehicle's maximal velocity for each point in time. 

 To generate the initial population for \nsga, we use Latin Hypercube Sampling~\citep{LHS}, a quasi-random sampling strategy, to initiate the search with a diverse population. We set the mutation rate to 1/3, and the crossover rate to 0.6, following existing guidelines \citep{Arcuri11parameter}.
We set the search budget for both NSGA-II and RS to 2,000 evaluations. For NSGA-II, we use the population size of 40 and set the number of generations to 50, as preliminary runs have shown, more generations do not significantly improve the coverage score evaluated with CID.

\textbf{Test oracles for AVP.} Similar to the MNIST case study, we define alternative test oracle functions for the AVP case study. Specifically, we use the following test oracles with increasing level of restriction: 
\setlength{\abovedisplayskip}{12pt} 
\setlength{\belowdisplayskip}{12pt} 
\begin{align*}
      O_{Large} (x):  f_1(x) < -0.7 \wedge f_2(x) < -1.\\
      O_{Medium} (x):  f_1(x) < -0.8 \wedge f_2(x) < -2.\\
      O_{Small} (x):  f_1(x) < -0.9 \wedge f_2(x) < -2.
\end{align*}

The oracle function \cmid considers a test case as failing when the ego vehicle is closer to the pedestrian and has  a higher speed compared to the oracle function \clarge. The test oracle \csmall constrains the conditions for failure even more, by requiring, compared to the two other functions, the highest maximal speed of the ego vehicle at the time of the minimal distance to the pedestrian.

\textbf{Diversified \nsga (\nsgad).} To answer RQ2, first we extend \nsga with an additional fitness function that evaluates the Euclidean distance of a test input to individuals found in previous generations (short, called \textit{Archive}).
Second, we apply a repopulation operator, which replaces in each generation the most dominated individuals by randomly sampled candidates. By maximizing the value of the additional fitness function and by employing the repopulation operator, we want to make the search identify more diverse test cases and therefore achieving a better coverage of failure-revealing tests. A similar approach has been used in DeepJanus by Riccio and Tonella~\citep{Riccio2020DeepJanus}. We do not directly apply the DeepJanus algorithm in our experiments since its main goal is to identify the borders of failure-revealing regions. However, we note that the diversity fitness function and the repopulation operator we use in our study are similar to those used by DeepJanus.
To configure the hyperparameters and processing technique for individuals to be included in the archive, we performed preliminary experiments. The archive uses a threshold to determine when a test input should be included. Before a solution can be added to the archive, the distance of the candidate to the closest individual in the archive is computed. A solution is only included in the archive if the distance value is higher than the threshold, promoting the storage of diverse inputs in the archive. As suggested by Riccio and Tonella~\citep{Riccio2020DeepJanus}, we selected the archive threshold use case specifically and compared the coverage performance when using the average versus the minimal Euclidean distance between failures, averaged over the results from all oracle definitions. Based on the results, we selected the minimal Euclidean distance between failing test inputs for both case studies, which was 0.29 for the AVP case study and 1.77 for MNIST.
Regarding the processing of individuals to be included into the archive, we compared a \textit{sequential processing}, where each individually is independently added into the archive, and \textit{population-based processing}.  In the sequential approach, adding an individual from the population to the archive can affect the distance computation for another individual in the population. In contrast, in a population-based processing first the distance is calculated for all individuals from a population to archived individuals. Based on whether the distance exceeds the threshold all allowed individuals are added together to the archive. In our preliminary experiments, sequential processing could outperform population based processing, why we selected to process sequentially individuals in our study.

\textbf{Experiments.} 
We repeat the execution of \nsga,  \mopso and RS 10 times to account for their randomness.
The AVP system is executed using the Prescan simulator \citep{Prescan} and configured with the default sampling rate of 100Hz. For the MNIST case study we execute \nsga, \mopso and 
RS with 20 different digits from the MNIST dataset depicting a 5 to ensure the generalization of our results. Before using a digit from the MNIST dataset for the study, we verify that it is correctly classified.

\textbf{Metrics.} To compute \metric{}, we need to develop a reference set, which approximates the actual set of failure-revealing test inputs, i.e., the DOI, for each case study.  The reference set is computed independently from the solution set of a search approach (test set).

For the AVP case study, we use two methods to generate the reference set: grid sampling (GS) and furthest point sampling (FPS), as explained in Section~\ref{sec:metrics_refset}. For GS, we consider three different resolutions, 10, 20 and 25 samples for each axis, generating in total 1,000, 8,000 and 15,625 samples. Executing 15,625 samples took about 43 hours. We did not consider resolutions higher then 25 samples per axis, since the CID value stabilizes when 25 samples are reached. For the FPS, we consider 1,000 and 8,000 samples, and did not consider 15,625 samples given the constrained time budget, because computing new samples in FPS is time intensive as the implementation of FPS is based on the generation of Voronoi cells. However, the relative comparison of different algorithms for both resolutions 1000 and 8000 remains the same.
Note, that we cannot verify that our reference sets are optimal as defined in Section~\ref{sec:metrics} since the sizes of some regions of the DOI may be less than the maximum distance between the adjacent points in our reference sets.
However, we can show the robustness and consistency of our results by considering different sampling methods, i.e., GS and FPS, and different sampling sizes, i.e., 1,000, 8,000, and 15,625.

For the MNIST case study, we use GS with a resolution of 10 samples per search dimension. Preliminary experiments demonstrate that using higher resolutions for GS or varying sampling methods does not significantly affect \metric{} results for MNIST. 





\begin{table}[t]
    \captionsetup{width=\textwidth}
    \centering
    \caption{Hyperparameter configuration for the search algorithms \nsga, \nsgad and 
\mopso 
for the case studies AVP and MNIST.}
    \begin{tabular}{l r r}
    \toprule
   Parameter & MNIST & AVP \\
    \midrule 
    \textbf{\nsga} & & \\
         population size & 20  &  40 \\
         mutation rate & 1/3  & 1/3 \\
         crossover rate & 0.6  &  0.6 \\
            \midrule 

       \textbf{\nsgad} & & \\
        population size & 20  &  40 \\
         mutation rate & 1/3  &  1/3 \\
         crossover rate & 0.6  &  0.6 \\  
         archive threshold  & 1.77  &  0.29 \\
         \midrule 
         \textbf{\mopso} & & \\
        swarm size & 20  &  40 \\
        mutation rate & 1/3 & 1/3 \\
        archive size & 20  &  40 \\
        boundary strategy & reflection & reflection \\
        min inertia weight & 0.1 & 0.1 \\
        weight update strategy & dynamic & dynamic \\
\bottomrule
    \end{tabular}
    \label{tab:config_algos}
\end{table}

\subsection{Experimental Results}
\label{sec:results}

\begin{figure}[t]
    \centering
    \includegraphics[scale=0.35]{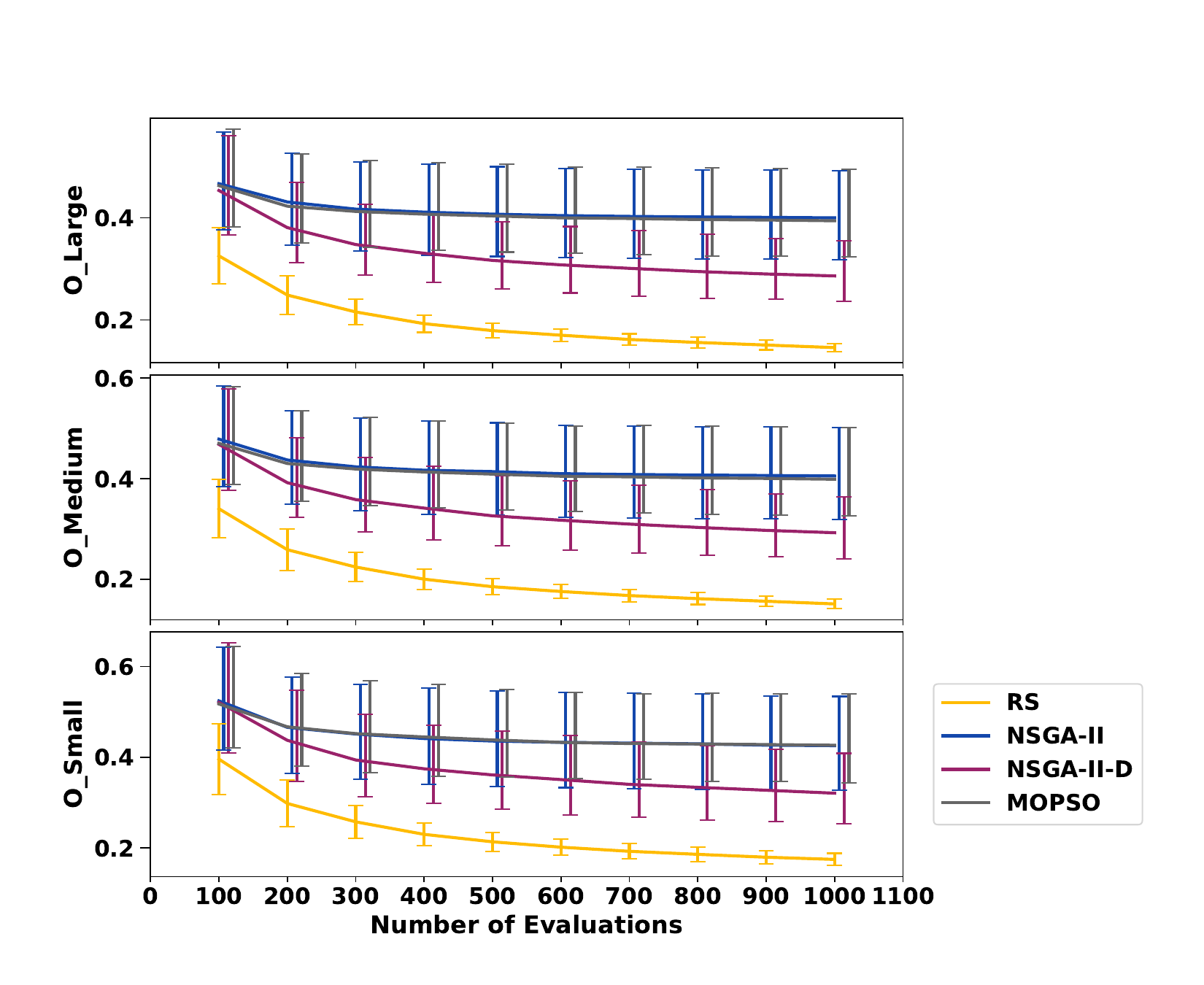}
    \caption{MNIST Case Study. The average and standard deviations of CID values obtained from 10 runs of RS, NSGA-II, NSGA-II-D and \mopso for the three test oracle functions \clarge, \cmid, and \csmall of MNIST. The \metric{} values are plotted after every 100 evaluations. The reference set is computed using 15,625 uniformly generated test inputs using Grid Sampling.}
    \label{fig:mnistcompare}
\end{figure}

\begin{figure}[t]
    \centering
    \includegraphics[scale=0.35]{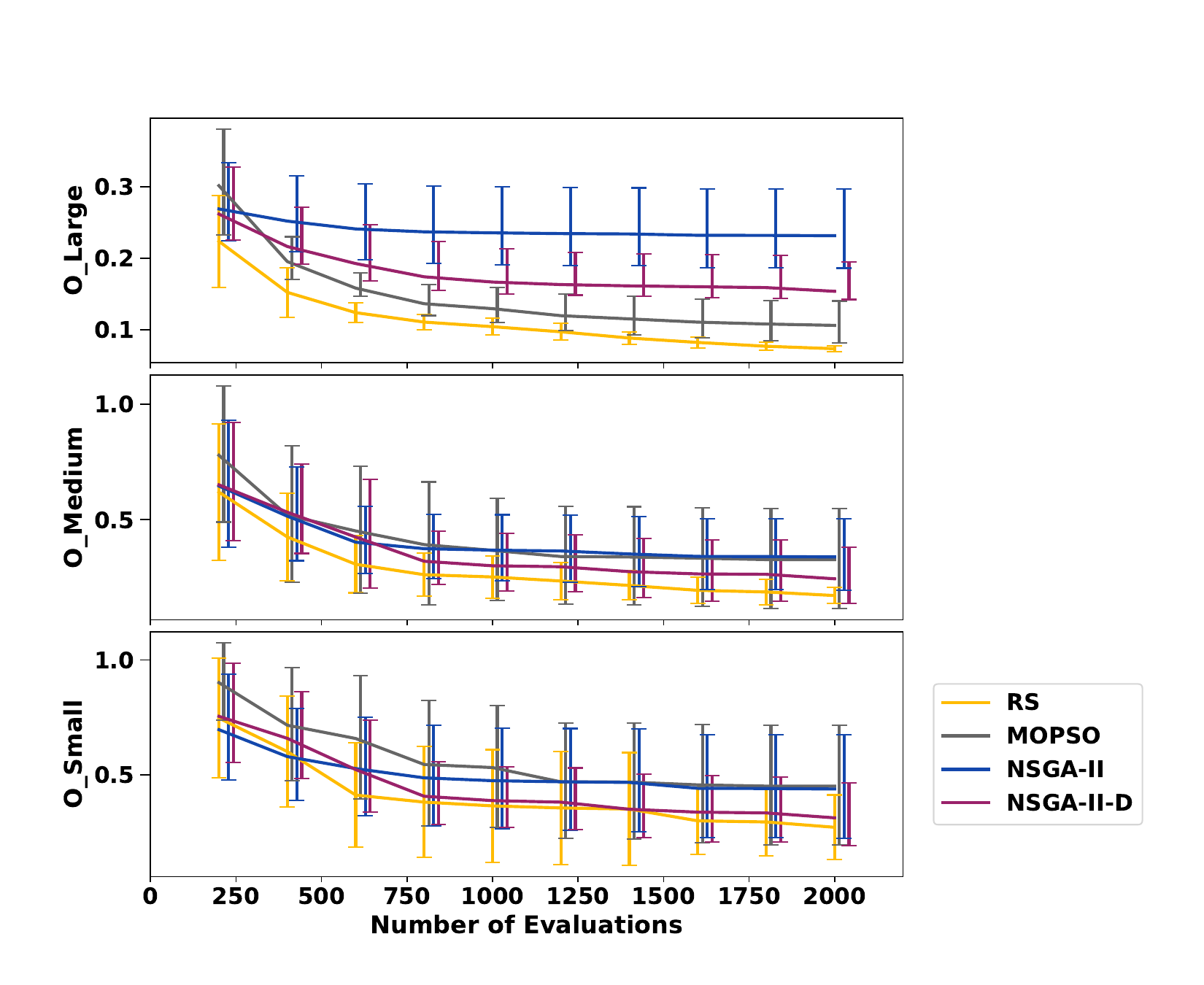}
    \caption{AVP Case Study. The average and standard deviations of CID values obtained from 10 runs of RS, \nsga, \nsgad and \mopso for the three test oracle functions \clarge, \cmid, and \csmall of AVP. The \metric{} values are plotted after every 250 evaluations. The reference set is computed using 15,625 uniformly generated test inputs using Grid Sampling.} 
    \label{fig:avpcompare}
\end{figure}

Figures~\ref{fig:mnistcompare} and \ref{fig:avpcompare} show the \metric{} results obtained by applying RS, \nsga, and \nsgad and \mopso to the MNIST and AVP case studies. For each system, the \metric{} results are computed with respect to their respective test oracle definitions that correspond to different DOI sizes. The diagrams in Figures~\ref{fig:mnistcompare} and \ref{fig:avpcompare} illustrate the average and standard deviations of the \metric{} values derived from 10 runs of each algorithm after every 100 and 250 evaluations, respectively.

The \metric{} values in the diagrams in Figures~\ref{fig:mnistcompare} and \ref{fig:avpcompare} are calculated using reference sets generated by grid sampling for each case study and each test oracle definition. We use the same reference set for each combination of algorithm, case study, and test oracle definition. However, the number of failures in the reference set depends on the oracle definition.

We note that due to our interest in coverage, we choose the set $A$, i.e., the solution set, used to compute CID as the set of all failures found by each testing algorithm over all its iterations. Specifically, for each of the \nsga, \nsgad, and \mopso algorithms, the set 
$A$ is the set of \emph{all} failures found over all generations and not just the failures in the last generation. Similarly, for random search, set $A$ is the set of \emph{all} failures found by the random search.

The AVP's reference sets contain 15,625 samples, while the MNIST's reference sets comprise 1,000 samples. In the remainder of this section, we discuss the results for RQ1 and RQ2 using the results in Figures~\ref{fig:mnistcompare} and \ref{fig:avpcompare}. 

\subsubsection{RQ1 Results}
\textbf{Comparing \metric{} results.} 
The MNIST Case Study results in Figure~\ref{fig:mnistcompare} show that RS consistently yields lower, i.e., better, average and lower standard deviations for \metric{} values than \nsga and \mopso over time across all test oracles. Moreover, although \metric{} values for \nsga, \mopso and RS decrease over time, the rate of reduction is more significant for RS compared to \nsga and \mopso.

Further, the results in Figure~\ref{fig:mnistcompare} show that while the \metric{} values for RS, \nsga and \mopso are similar for the \cmid and \clarge test oracles, they are higher for the \csmall test oracle. This indicates that covering the DOI regions for \csmall, the strictest test oracle definition, poses a greater challenge for both algorithms compared to \cmid and \clarge. Finally, for all three cases, the \metric{} values of \ras, \nsga and \mopso stabilize in less than 1000 evaluations. 



Similarly, the results presented in Figure~\ref{fig:avpcompare} demonstrate that for AVP, the \metric{} values achieved by RS consistently outperform those obtained by \nsga across all test oracles and after at least 300 evaluations. Further, the average of \metric{} values and the variations of \metric{} values obtained by RS decrease at a higher rate compared to those obtained by \nsga and \mopso. However, \mopso can outperform \nsga for the less constrained oracle definition $O_{Large}$ in \metric{} values. For the $O_{Medium}$ and $O_{Small}$ oracles, there is no significant difference in $\metric{}$ values between \mopso and \nsga.

In each of the three diagrams, the \metric{} values for RS, \nsga and \mopso stabilize within 2000 evaluations, suggesting that executing the algorithms longer is unlikely to change the comparison of the \metric{} values.

In addition, two observations can be drawn from Figure~\ref{fig:avpcompare}: (1) For both \nsga and \ras, stricter oracle definitions, corresponding to smaller DOIs and fewer failures in a large search space, result in higher metric values, indicating that covering failures is more challenging when there are fewer failures. (2) The difference in \metric{} values between \nsga and \ras decreases with a stricter oracle function. This is expected; as DOIs become smaller and failures harder to detect, \ras may become less effective in covering and exploring failures, behaving more similarly to \nsga and \mopso.  Nevertheless, even for our strictest notion of a test oracle, \nsga and \mopso cannot outperform \ras in covering DOI for both case studies.

\begin{table}
    \captionsetup{width=\textwidth}
\caption{(Reference Set Variation) The average and standard deviations of \metric{} values obtained from 10 runs of RS, \nsga, \nsgad, \mopso for the AVP system with the \clarge, $O_{Medium}$ and \csmall test oracles for different reference sets and sizes. The \metric{} values are computed using reference sets obtained by grid sampling (GS) and furthest point sampling (FPS). The number of test inputs generated for the reference set is denoted in brackets. Best \metric{} value is marked in bold.} \label{tab:comparison_refset}
\centering
         \begin{tabular}{l cc cc cc cc}
        \toprule
         & \multicolumn{2}{c}{\textbf{\nsga}} 
         & \multicolumn{2}{c}{\textbf{\nsgad}}
         & \multicolumn{2}{c}{\textbf{RS}} 
         & \multicolumn{2}{c}{\textbf{\mopso}}\\ 
        \cmidrule(r){2-3}\cmidrule(l){4-5}\cmidrule(l){6-7}\cmidrule(l){8-9}
         & Avg & Std    &  Avg & Std      &  Avg & Std    &  Avg & Std    \\
         \midrule
         $O_{Large}$  &   & &  & &  & &  & \\
         \midrule
        \textbf{GS (1000)} & 0.228 & 0.062 &  0.162 & 0.025  & \textbf{0.076} & 0.005 & 0.114 & 0.030\\
         \textbf{FPS (1000)} & 0.253 & 0.071 & 0.169 & 0.029 & \textbf{0.076} & 0.008 & 0.118 & 0.026  \\
         \textbf{GS (8000)} & 0.229 & 0.059 & 0.160 & 0.025 &  \textbf{0.071} & 0.004 & 0.107 & 0.029 \\
         \textbf{FPS (8000)} & 0.227 & 0.062 & 0.156 & 0.027 & \textbf{0.069} & 0.003 &  0.105 & 0.030 \\ 
           \textbf{GS (15625)} & 0.231  &0.055 &  0.153 & 0.026 & \textbf{0.073} & 0.004  & 0.106 & 0.029\\\midrule
                       $O_{Medium}$ &  & & & & &   & &  \\
         \midrule

        \textbf{GS (1000)} & 0.445 & 0.254 & 0.355 & 0.127 & \textbf{0.259} & 0.064& 0.489 & 0.338 \\
         \textbf{FPS (1000)} & 0.594 & 0.439  & 0.500 & 0.321 & \textbf{0.349} & 0.143 & 0.670 & 0.630 \\
          \textbf{GS (8000)} & 0.352 & 0.147  & 0.280 & 0.134 & \textbf{0.182} & 0.039 & 0.340 & 0.219 \\
          \textbf{FPS (8000)} & 0.348 & 0.152  & 0.285 & 0.151 & \textbf{0.184} & 0.040 & 0.346 & 0.213   \\
                     \textbf{GS (15625)} & 0.338 & 0.155 & 0.243 & 0.122 &  \textbf{0.171} & 0.035 & 0.326 & 0.216 \\
          \midrule
                   $O_{Small}$ &   & &  & &  & &  & \\
         \midrule
        \textbf{GS (1000)} & 0.532 & 0.210 & 0.447 & 0.133 & \textbf{0.382} & 0.167 & 0.608 & 0.313 \\
         \textbf{FPS (1000)} & 0.691 & 0.243  & 0.592 & 0.330 & \textbf{0.497} & 0.297 & 0.653 & 0.374 \\
                  \textbf{GS (8000)} & 0.437 & 0.224  & 0.319 & 0.155 & \textbf{0.259} & 0.143 & 0.442 & 0.268 \\
          \textbf{FPS (8000)} & 0.454 & 0.230  & 0.337 & 0.171 & \textbf{0.279} & 0.166 & 0.459 & 0.253   \\
                     \textbf{GS (15625)} & 0.438 & 0.225 & 0.313 & 0.136 &  \textbf{0.272} & 0.140 & 0.451 & 0.260 \\
          \bottomrule
        \end{tabular}
\end{table}

\textbf{Sampling methods and sampling resolutions.}
As mentioned in Section~\ref{sec:setup}, the \metric{} results shown in Figures~\ref{fig:mnistcompare} and \ref{fig:avpcompare} are derived from specific reference sets generated using grid sampling and with a specific sampling resolution. One might wonder how altering the sampling method or resolution would affect the comparison of \metric{} values between RS, \nsga and \mopso. 
Table~\ref{tab:comparison_refset} shows the average and standard deviation of \metric{} values for the AVP system with the \clarge test oracle, calculated using the three sampling methods discussed in Section~\ref{sec:metrics_refset},  grid sampling (GS) and furthest point sampling (FPS) reference sets with 1,000 and 8,000 samples. As the table shows, even if we use reference sets obtained from different sampling methods and with different sample sizes, RS still fares considerably better than \nsga, \nsgad and \mopso. This indicates the robustness of our results when we use different sampling techniques and different number of samples to generate the reference sets to compute \metric{} values. Note that the results in Table~\ref{tab:comparison_refset} for 1,000 and 8,000 samples for both GS and FPS for $O_{Large}$ are close, indicating that the reference set of size 1,000 already provides a good approximation of the DOI.

For AVP the underlying number of failures in the reference sets for Grid Sampling with the highest resolution of in total 15,625 samples across all oracle definitions is presented in Section~\ref{sec:lessons} in Table~\ref{tab:num-failures}.









\textbf{Statistical Tests.} We use the non-parametric pairwise Wilcoxon
rank sum test and the Vargha-Delaney’s $\hat{A}_{12}$ effect size to compare the \metric{} results in Figures~\ref{fig:mnistcompare} and \ref{fig:avpcompare} between random search and the Pareto-based approaches. The level of significance has been set to 0.05. The results are presented in the first two rows of Table \ref{tab:stats-tests}. We adopt the following standard classification for effect size values: an effect size $e$ is small, when $0.56 \leq e \leq 0.64$ or  $0.34 \leq e \leq 0.44$, is medium when $0.64 < e \leq 0.71$  or $0.29 \leq e   \leq  0.34$ and large when it is higher then $0.71$ or lower then $0.29$. Otherwise the effect size is negligible.

The statistical test results show that for MNIST and AVP with  \clarge and \cmid test oracles, RS yields CID values that are significantly better than those generated by \nsga 
and \mopso with large effect sizes. However, for the \csmall test oracle of AVP, the difference between RS and \nsga is not statistically significant. This is consistent with the results in Figure~\ref{fig:avpcompare} for the AVP with the \csmall and the \cmid test oracles. 

For the less constrained test oracles, \clarge and \cmid, RS significantly outperforms \nsga and \mopso in covering the failure-revealing tests, while for the more constrained \csmall test oracle the difference between the algorithms \nsga, \mopso and \rs diminishes and neither covers the DOI better than the other.

For completeness, we have provided in Table~\ref{tab:stats-heuristics} also the statistical-test results for the comparison of CID results between \nsga and \mopso in the second row. We can see that for the AVP case study only for oracle \clarge the difference between \nsga and \mopso is statistically significant with a large effect size what is consistent with the result in Figure~\ref{fig:avpcompare}. For MNIST however, there is not statistical significance what is in line with the results in Figure~\ref{fig:mnistcompare}.

\begin{table}[t]
\captionsetup{width=\textwidth}
\centering
\caption{Statistical-test results for comparison between RS and Pareto-based approaches (Wilcoxon and Vargha-Delaney). The first and the second rows compare the CID values obtained by RS and NSGA-II in the last iterations in Figures~\ref{fig:mnistcompare} and~\ref{fig:avpcompare}. The third and fourth rows compare the CID values obtained by RS and NSGA-II-D in the last iterations of Figures~\ref{fig:mnistcompare} and~\ref{fig:avpcompare}. The fifth and sixth rows compare the CID values obtained by RS and \mopso in the last iterations of Figures~\ref{fig:mnistcompare} and~\ref{fig:avpcompare}. The letter L represents a \textit{large} effect size magnitude.}\label{tab:stats-tests}
\begin{tabular}{m{2.7cm} c l c c c}
\toprule
 & \textbf{Case Study} & \textbf{Measure} & \bclarge & \bcmid & \bcsmall \\ 
\hline
\textbf{RS vs. \nsga} & AVP & p\_value & \textbf{0.0019} & \textbf{0.0019} & 0.0839 \\
 &  & effect\_size & \textbf{0 (L)} & \textbf{0.09 (L)} & 0.23 (L) \\
\bottomrule
  & MNIST & p\_value & $\boldsymbol{\sim 0}$ & $\boldsymbol{\sim 0}$ & $\boldsymbol{\sim 0}$ \\
  & & effect\_size & \textbf{0.02 (L)} & \textbf{0.02 (L)} & \textbf{0.02 (L)} \\
\hline
\textbf{RS vs. \nsgad} & AVP & p\_value & \textbf{0.0019} & 0.083 & 0.4921 \\
 & & effect\_size & \textbf{0 (L)} & 0.25 (M) & 0.39 (S) \\
\bottomrule
   & MNIST & p\_value & $\boldsymbol{\sim 0}$ & $\boldsymbol{\sim 0}$ & $\boldsymbol{\sim 0}$ \\
    & & effect\_size & \textbf{0.07 (L)} & \textbf{0.08 (L)} & \textbf{0.08 (L)} \\ 
\hline
\textbf{RS vs. \mopso} & AVP & p\_value & \textbf{0.0019} & \textbf{0.0097} & \textbf{0.0488} \\
 & & effect\_size & \textbf{0 (L)} & \textbf{0.13 (L)} & \textbf{0.23 (L)} \\
\bottomrule
   & MNIST & p\_value & $\boldsymbol{\sim 0}$ & $\boldsymbol{\sim 0}$& $\boldsymbol{\sim 0}$ \\
    & & effect\_size & \textbf{0.02 (L)} & \textbf{0.02 (L)} & \textbf{0.02 (L)} \\
\bottomrule
\end{tabular}
\end{table}

\begin{table}[t]
\captionsetup{width=\textwidth}
\centering
\caption{Statistical-test results for Pareto-based approaches (Wilcoxon and Vargha-Delaney). The first and the second rows compare the CID values obtained by NSGA-II and NSGA-II-D in the last iterations in Figures~\ref{fig:mnistcompare} and~\ref{fig:avpcompare}. The third and fourth rows compare the CID values obtained by NSGA-II and \mopso in the last iterations of Figures~\ref{fig:mnistcompare} and~\ref{fig:avpcompare}. The fifth and sixth rows compare the CID values obtained by \mopso and \nsgad in the last iterations of Figures~\ref{fig:mnistcompare} and~\ref{fig:avpcompare}. The letter L represents a \textit{large} effect size magnitude.}
\label{tab:stats-heuristics}
\begin{tabular}{m{2.5cm} c l c c c}
\toprule
 & \textbf{Case Study} & \textbf{Measure} & \bclarge & \bcmid & \bcsmall \\ 
\midrule
\textbf{NSGA-II vs. NSGA-II-D} & AVP & p\_value & \textbf{0.0019} & 0.3222 & 0.4316 \\ 
 & & effect\_size & \textbf{0.93 (L)} & 0.63 (S) & 0.65 (M) \\ 
\hline
 & MNIST & p\_value & $\boldsymbol{\sim 0}$ & $\boldsymbol{\sim 0}$ & $\boldsymbol{\sim 0}$ \\ 
 & & effect\_size & \textbf{0.82 (L)} & \textbf{0.82 (L)} & \textbf{0.80} (L) \\ 
\hline
\textbf{NSGA-II vs. OMOPSO} & AVP & p\_value & \textbf{0.0019} & 0.6953 & 0.9218 \\ 
 & & effect\_size & \textbf{1.0 (L)} & 0.59 (S) & 0.545 (S) \\ 
\hline
 & MNIST & p\_value & 0.9854 & 1.0  & 1.0\\ 
 & & effect\_size & 0.50 (N) & 0.50 (N) & 0.48(N) \\ 
\hline
\textbf{OMOPSO vs. NSGA-II-D} & AVP & p\_value & \textbf{0.0019} & 0.6953 & 0.3222 \\ 
 & & effect\_size & \textbf{0.11 (L)} & 0.60 (S) & 0.67 (M) \\ 
\hline
 & MNIST & p\_value & $\boldsymbol{\sim 0}$ & $\boldsymbol{\sim 0}$ &  $\boldsymbol{\sim 0}$ \\ 
 & & effect\_size & \textbf{0.83 (L)} & \textbf{0.82 (L)} & \textbf{0.82 (L)} \\ 
\bottomrule
\end{tabular}
\end{table}
\begin{tcolorbox}

The answer to RQ1 is that \nsga and \mopso perform less effectively than na\"ive random testing in covering failure-revealing test inputs for the AVP and MNIST case studies. We confirm the robustness of the sampling method and the sampling resolution used to compute \metric{} values in our experiments, as our results remain unchanged with a more fine-grained sampling resolution and the use of different sampling methods. We, further, demonstrate that our comparison remains consistent across 11 of the 12 alternative test oracle definitions for our two case studies, yielding statistically significant differences with large effect sizes. Overall, our results across both case studies and the 12 alternative test oracle function definitions for these case studies show that testing techniques based on the Pareto optimization algorithms, \nsga and \mopso, cannot outperform RS in covering failures.
\end{tcolorbox}

\subsubsection{RQ2 Results}
\label{sec:results_rq2}


The results in Figures~\ref{fig:mnistcompare} and~\ref{fig:avpcompare} show that the \metric{} values obtained by \nsgad are always better than those obtained by \nsga over time for both case studies and all test oracle definitions. However, compared to RS, \nsgad achieves consistently worse \metric{} values for all test oracle definitions of the MNIST case study and the \clarge test oracle of the AVP case study. For the \cmid and \csmall test oracles of the AVP case study, RS outperforms \nsgad after 500 and 1500 evaluations respectively.

\textbf{Statistical Tests.} The statistical-test results in Table~\ref{tab:stats-tests} show that RS yields CID values that are significantly better than those generated by \nsgad with large effect sizes for all test oracle definitions of MNIST and the \clarge test oracle of AVP. Similar to the comparison between RS and \nsga, the difference between RS and \nsgad for the \csmall test oracle of AVP is not statistically significant. For \csmall , RS does not outperform \nsgad, nor does \nsgad outperform RS. This indicates that even when \nsga is enhanced with a diversity-focused fitness function, it still does not outperform RS in terms of failure coverage.

The statistical test results comparing \nsgad with the Pareto algorithms \nsga and \mopso indicate a statistically significant difference between \mopso and \nsgad for the MNIST case study across all oracles. For the AVP case study, this difference is significant only for the \clarge oracle. The difference between \nsgad and \nsga is statistically significant for the AVP case study only for the \clarge oracle, and for the MNIST case study for all oracles. These findings are consistent with the results presented in Figure~\ref{fig:avpcompare} and Figure~\ref{fig:mnistcompare}.



\begin{tcolorbox}
The answer to RQ2 is that while a diversified search using Pareto-based optimization can improve the coverage of failure-revealing tests, the search still remains less effective than na\"ive random testing in covering failure-revealing test inputs. In particular, on average, \nsgad reduces the differences in \metric{} values between \nsga and RS by up to 60.6\% for the AVP case study, and by up to 43.7\% for the MNIST case study. However, the metric values produced by \nsgad remain significantly worse than those generated by RS across 4 of the six alternative test oracle definitions for our two case studies, yielding statistically significant differences with large effect sizes. For the remaining two oracle definitions the metric values are not better compared to RS.
\end{tcolorbox}

\hfill

\section{Threats to Validity}
\label{sec:usage}
The most important threats concerning the validity of our experiments are related to the internal, external and construct validity. 


To mitigate \textit{internal validity} risks, which refer to confounding factors, we used identical search spaces and search configurations for all the algorithms -- RS, \nsga, \nsgad and \mopso -- in each of our case studies: AVP and MNIST. For MNIST, the objective is to create test inputs that challenge classifiers while remaining valid and preserving the digit's original label. ``Valid" implies that the digits are still recognizable by humans, and ``label preserving" ensures that the image's label remains unchanged after mutation; for instance, a mutated ``5" should not transform into a ``6". To address this, we have taken the following actions: (1)~We ensure the validity and label-preservation of individuals in the initial population for our search algorithms by enforcing a distance threshold, as suggested by the previous studies~\cite{Riccio2020DeepJanus}, on the distances between the individuals and initial seed. (2)~We select images from the MNIST dataset as seeds only if they can be correctly classified by the classifier under test. (3)~We use a small displacement interval in the mutation operator to reduce the chances of altering the label or shape of digit images through perturbations. (4)~Finally, we identify out-of-bound displacements in images during mutation and adjust the mutated images by reversing the displacement, bringing it back within the image boundary.

To mitigate the bias associated with choosing suboptimal hyperparameters for experiments using \mopso and \nsgad, we conducted preliminary experiments to identify optimal parameter configuration, as outlined in Section~\ref{sec:setup}.



\textit{Construct validity} threats relate to the inappropriate use of metrics. To compare the coverage of the failure-revealing domain of test inputs for the studied search approaches, we have applied the \metric{} metric introduced in Section~\ref{sec:metrics_qi}.
To the best of our knowledge, there have been no other metrics in the literature that serve our purpose. We have outlined in Section~\ref{sec:related-work} related metrics and have positioned our metric.


\textit{External validity} is related to the generalizability of our results. We used two case studies from different domains: the AVP case study is an industrial ADAS, and the MNIST case study is a widely-used open-source system for classifying handwritten digits. The specification of AVP and the definition of its search space are provided by our industrial partner. Furthermore, we have used a widely-used, stable and high-fidelity simulator Prescan to simulate the ADAS. We apply search-based testing to the MNIST case study in a manner consistent with prior studies using the same case study~\cite{Zohdinasab21DeepHyperion, Riccio2020DeepJanus}. Regarding the choice of the Pareto-based algorithms, in our experiments, we considered two widely-used and different types of Pareto-based techniques, \nsga and \mopso. The diversity fitness function and the repopulation operator, both employed to improve \nsga, are adopted from recent research aimed at introducing diversity into \nsga~\cite{Riccio2020DeepJanus}. 

Further, we provide an argument in Section~\ref{sec:theoretical} based on the definition of Pareto-based optimization and which is independent of the system under test, regarding why Pareto-based search techniques are not suitable for the coverage of failure-revealing test inputs. The argument holds under the assumptions defined in Section~\ref{sec:preliminaries}.


\textit{Replicability.} To foster replicability, the implementation of the \metric{} metric, as well as results of the evaluation and conducted experimental runs with RS, \mopso, \nsga and diversified search with \nsga are available online \cite{replication-package}. The implementation of the AVP case study could not have been disclosed as it was developed by our industrial partner. The search approach and the analysis of the results have been implemented using the modular and open-source search-based testing framework OpenSBT \cite{sorokin2023opensbt}.

\section{Discussion}
\label{sec:lessons}
In this section, we offer further observations and  discuss the applications and limitations of our proposed metric, \metric{}. In addition, we 
compare \nsga, \nsgad, \mopso and RS in terms of their effectiveness and efficiency in finding failure instances. This comparison aims to better position our study within the context of prior research and complements the study presented in Section~\ref{sec:application}, which is primarily focused on comparing these algorithms with respect to the coverage of failure-inducing regions, as opposed to their ability to identify individual failures.

\begin{figure*}[t]
  \centering
  \begin{subfigure}[t]{0.25\textwidth}
       \includegraphics[scale=.3]{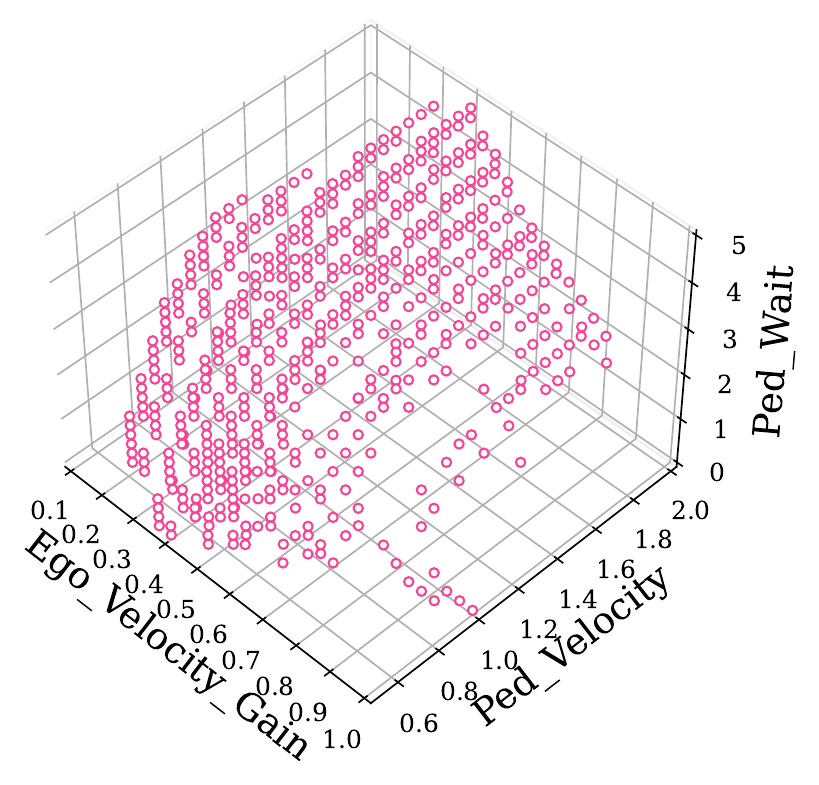}
    \caption{Reference set obtained by uniformly generating test data (15,625 samples, 558 failing).}
    \label{fig:3d_plot_gs_large}
  \end{subfigure}
  \hfill
  \begin{subfigure}[t]{0.25\textwidth}
    \includegraphics[scale=.30]{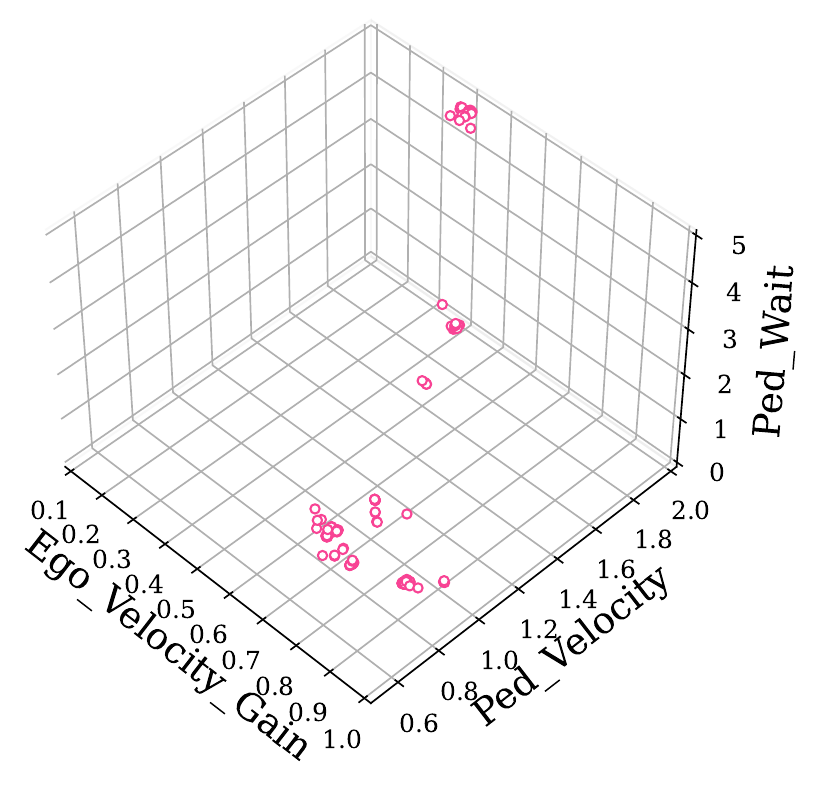}
     \caption{Failure-revealing solutions found by \nsga (429 solutions).}
    \label{fig:3d_plot_NSGA-II_large}
  \end{subfigure}
  \hfill
    \begin{subfigure}[t]{0.25\textwidth}
   \includegraphics[scale=.30]{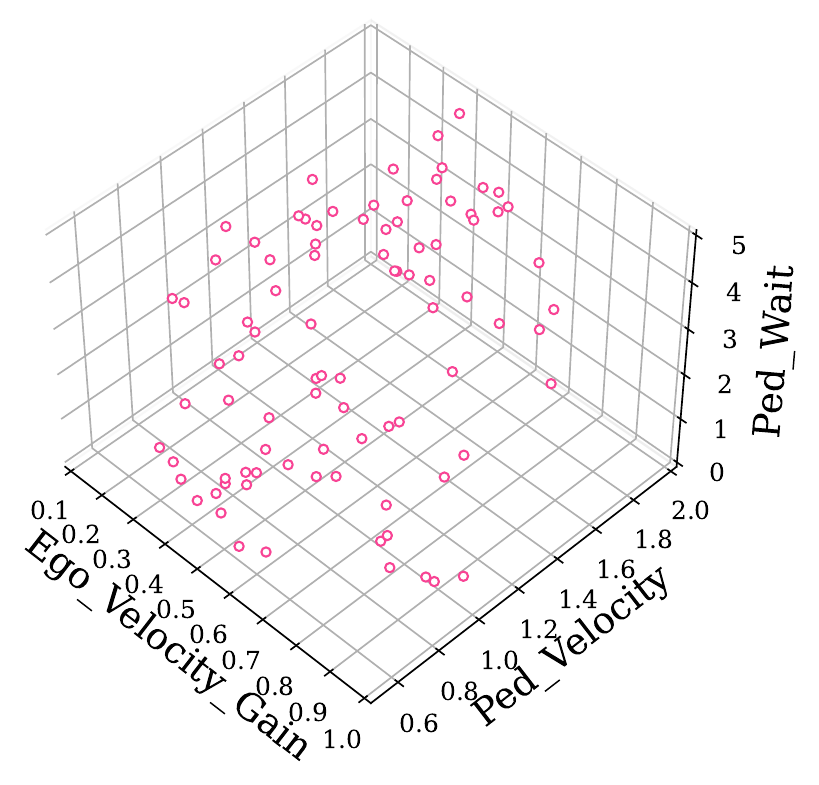}
   \caption{Failure-revealing solutions found by random search (89 solutions).}
    \label{fig:3d_plot_rs_large}
  \end{subfigure}
   \begin{subfigure}[t]{0.25\textwidth}
   \includegraphics[scale=.30]{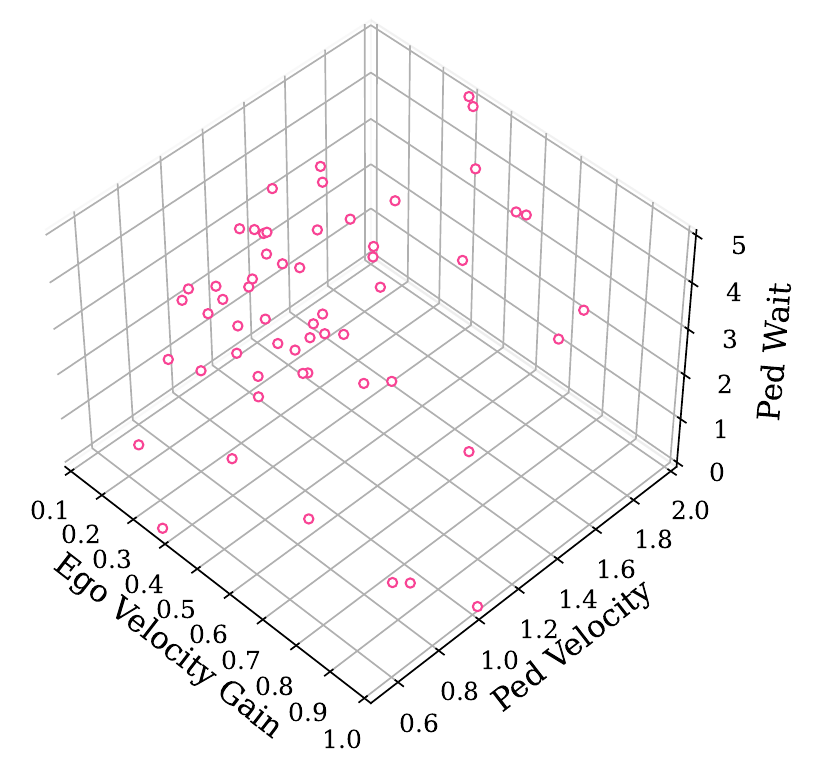}
   \caption{Failure-revealing solutions found by \mopso (57 solutions).}
    \label{fig:3d_plot_mopso_large}
  \end{subfigure}
  \qquad
   \begin{subfigure}[t]{0.25\textwidth}
   \includegraphics[scale=.30]{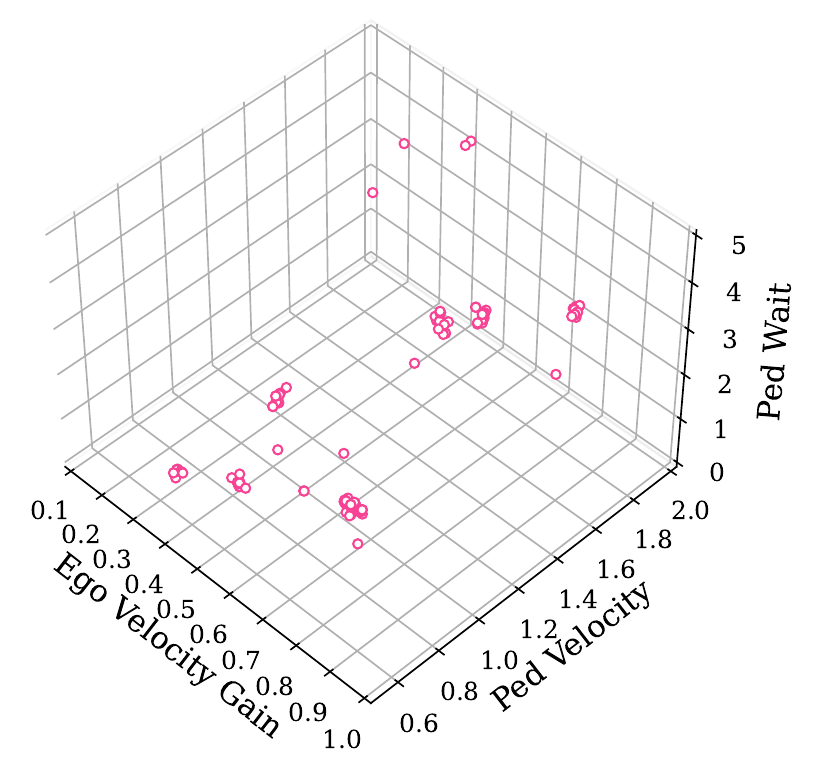}
   \caption{Failure-revealing solutions found by \nsgad (137 solutions).}
    \label{fig:3d_plot_nsga2d_large}
  \end{subfigure}
  \caption{Visualizing failure-revealing solutions  in the search space for AVP with the \clarge test oracle function. Plot (a): failure-revealing tests identified using grid sampling with 15,625 samples. This set serves as a reference set.  Plot (b) to (d) show failure-revealing tests found using one run of \nsga, random search, \mopso and \nsgad respectively. Although random search finds far fewer failure-revealing solutions than \nsga and \nsgad, it covers the DOI in Plot (a) more effectively and yields a lower \metric{}.}
   \label{fig:ds_results_clarge}
\end{figure*}

\textbf{Observation:} \textit{Why do the Pareto-based approaches \nsga, \nsgad and \mopso yield worse \metric{} values with considerably larger variations compared to RS?} To understand why \nsga, \nsgad and \mopso yield worse \metric{} values with considerably larger variations compared to RS, we plot the failure-revealing solutions obtained by \nsga, \nsgad, \mopso and RS, along with the reference sets computed for RQ1 to calculate \metric{} values. Recall that reference sets approximate the complete set of failure-revealing test inputs within the search space. Specifically, 
Figure~\ref{fig:3d_plot_gs_large} shows the reference set for AVP with the test oracle \clarge computed based on grid sampling with 15,625 samples as part of our RQ1 experiments, and 
Figure~\ref{fig:3d_plot_NSGA-II_large} to Figure~\ref{fig:3d_plot_nsga2d_large}, respectively, show the failing test inputs computed by one run of \nsga, RS, \mopso and \nsgad for AVP with the \clarge test oracle. The Figures contain all failures that have been identified over all evaluations of the run of their respective algorithm.

By comparing the failure-revealing test inputs in Figures~\ref{fig:3d_plot_NSGA-II_large} and \ref{fig:3d_plot_rs_large} with those from the reference set shown in Figure~\ref{fig:3d_plot_gs_large}, which locates 558 failures, one can see that although RS finds far fewer failure-revealing solutions than \nsga, 89 versus 429, RS covers the DOI in Plot (a) more effectively.

The failure-revealing tests identified by RS are more spread out and diverse, while those found by \nsga are grouped close together. Similarly, when comparing the failures found by \nsga and \nsgad in Figure~\ref{fig:3d_plot_nsga2d_large}, one can see that while the failures found by \nsgad are more scattered compared to \nsga, they are still clustered together and are not as spread as the solutions found by \ras.

\begin{figure*}[t]
  \centering
  \begin{subfigure}[t]{0.28\textwidth}
       \includegraphics[scale=.33]{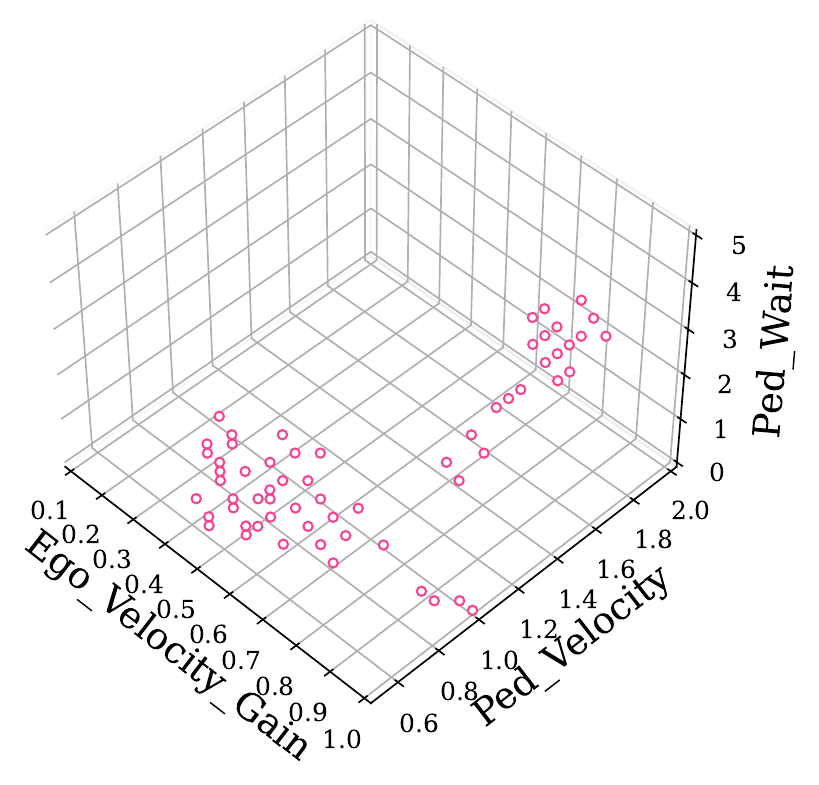}
    \caption{Reference set obtained by uniformly generated test data (15,625 samples, 62 failing).}
    \label{fig:3d_plot_gs_small}
  \end{subfigure}
  \hfill
  \begin{subfigure}[t]{0.28\textwidth}
    \includegraphics[scale=.33]{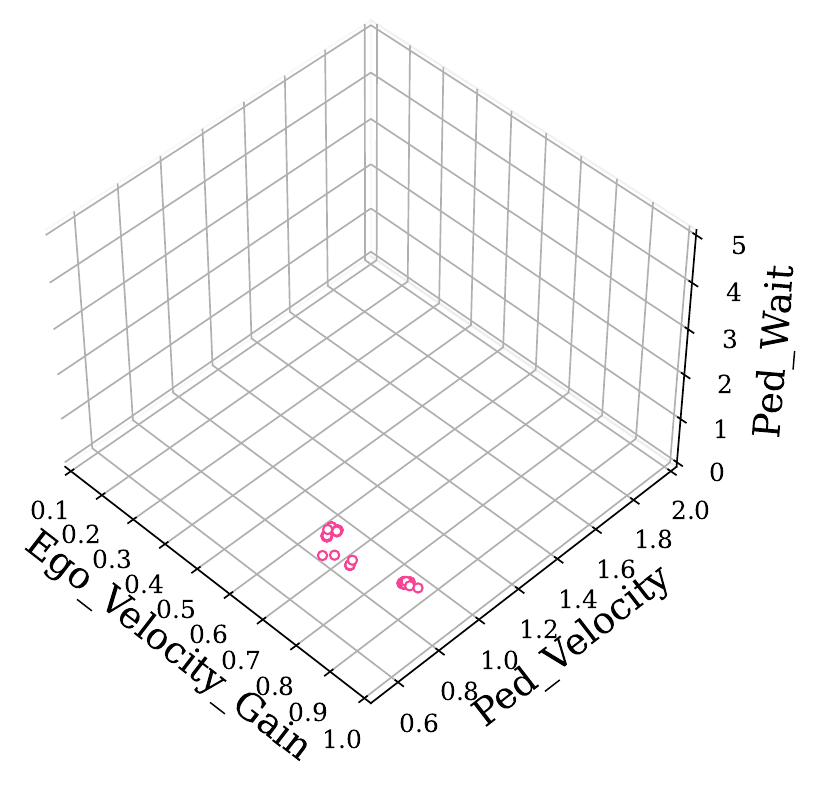}
     \caption{Failure-revealing solutions found by \nsga (111 solutions).}
    \label{fig:3d_plot_NSGA-II_small}
  \end{subfigure}
  \hfill
    \begin{subfigure}[t]{0.33\textwidth}
    \includegraphics[scale=.33]{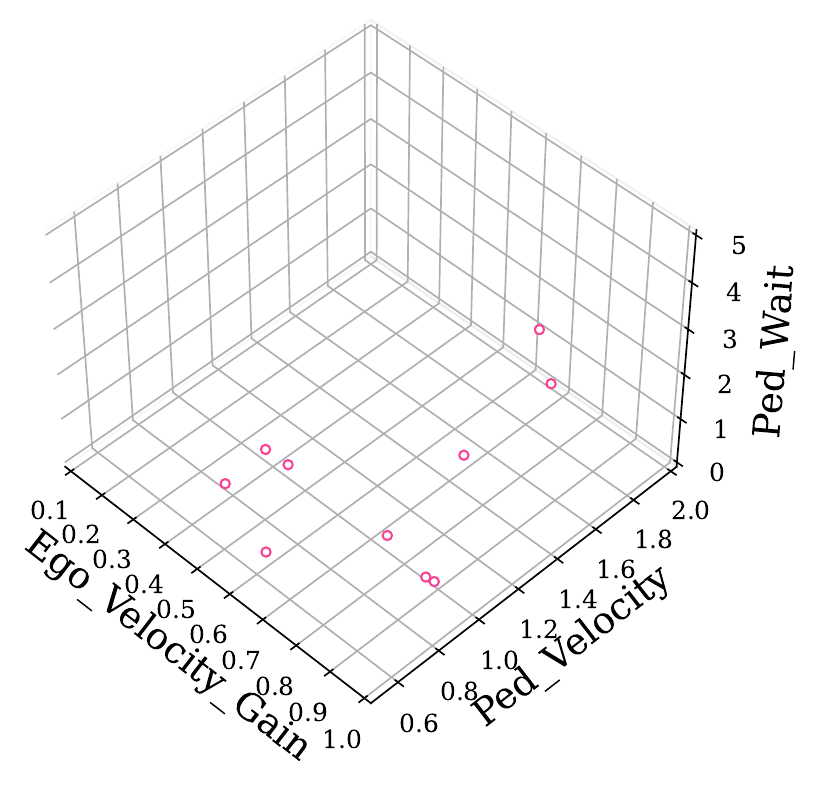}
   \caption{Failure-revealing solutions found by random search (10 solutions).}
    \label{fig:3d_plot_rs_small}
  \end{subfigure}
    \begin{subfigure}[t]{0.33\textwidth}
    \includegraphics[scale=.33]{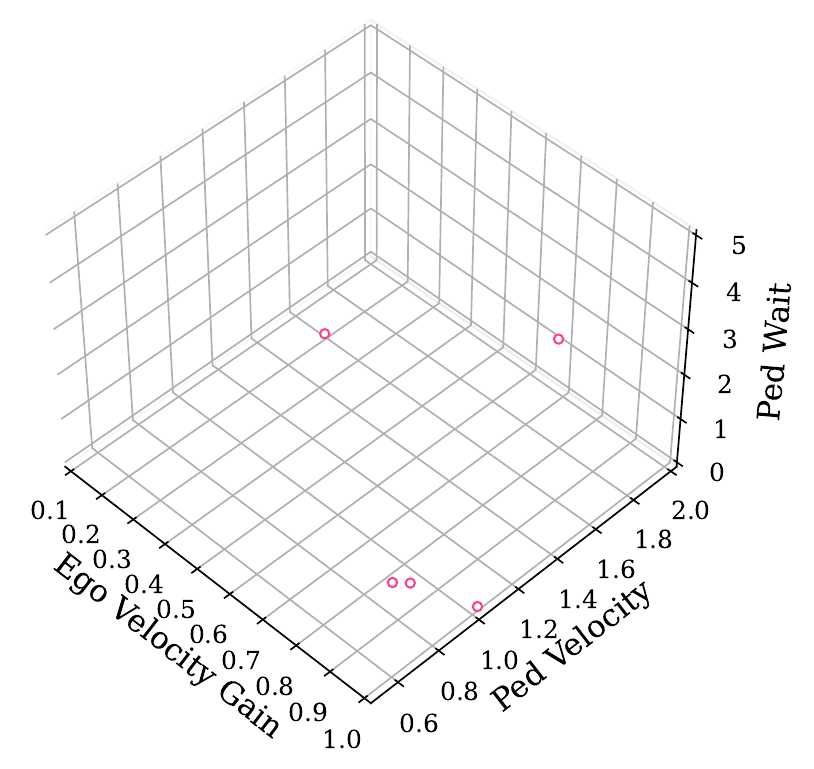}
   \caption{Failure-revealing solutions found by \mopso (5 solutions).}
    \label{fig:3d_plot_mopso_small}
  \end{subfigure}
   \qquad
   \begin{subfigure}[t]{0.33\textwidth}
    \includegraphics[scale=.33]{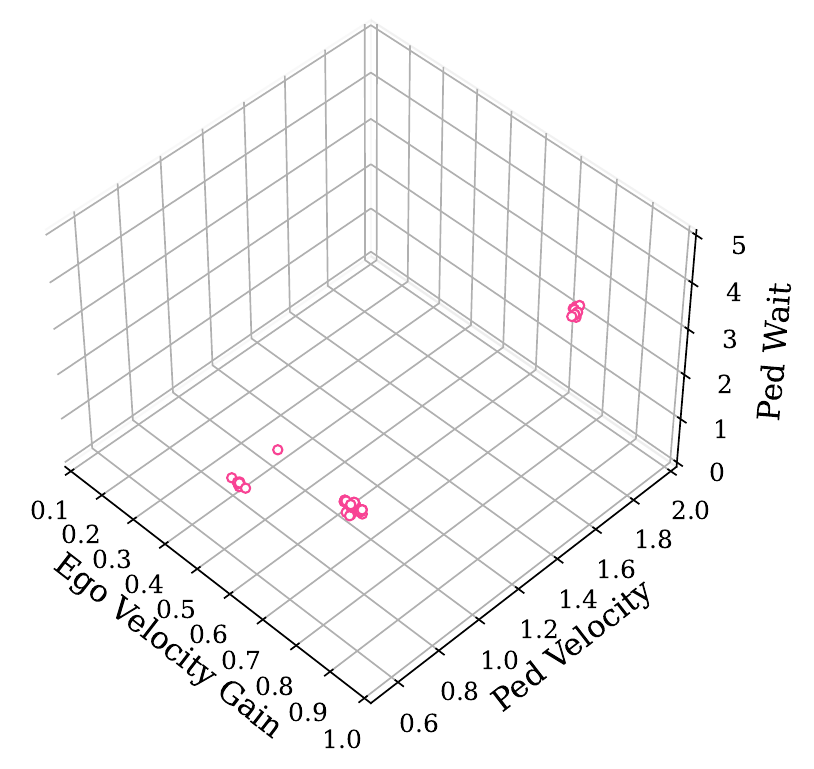}
   \caption{Failure-revealing solutions found by \nsgad (69 solutions).}
    \label{fig:3d_plot_nsga2d_small}
  \end{subfigure}
  \caption{Visualizing failure-revealing solutions for AVP in the search space for the most constrained oracle function \csmall.  Plot (a) shows the reference set obtained to approximate failures of the system under test. Plot (b) to (d) show failure-revealing tests found using one run of \nsga, random search, \mopso and \nsgad respectively. Although \nsga and \nsgad identify more failure-revealing test inputs than RS, RS achieves a better coverage of the DOI than \nsga.} 
    \label{fig:ds_results_csmall}
\end{figure*}

\mopso  identifies only 57 failing tests, but the failures as shown in Figure~\ref{fig:3d_plot_mopso_large} are more spread out compared to \nsga and \nsgad. However, we can see that larger regions of the test input space are left uncovered compared to \ras.
Similarly, we plot in Figure \ref{fig:ds_results_csmall}a the reference set for the AVP case study with the \csmall test oracle computed using the same sampling method and resolution as the one in Figure~\ref{fig:3d_plot_gs_large}. As shown in this figure, only 62 failing tests are identified in the reference set for  \csmall, which is significantly less then the number of failures in  the reference set  for the \clarge oracle.
Further, the failure-revealing solutions obtained by one run of \nsga and RS are shown in  Figure~\ref{fig:ds_results_csmall}b and  Figure~\ref{fig:ds_results_csmall}c, respectively.

RS identifies 10 failures, \nsga identifies 111 failures, \nsgad identifies 69 failing tests, while \mopso finds 5 failures. The failing test inputs found in one run of \mopso and \nsgad are shown in Figure~\ref{fig:ds_results_csmall}d and Figure ~\ref{fig:ds_results_csmall}e. The figures illustrate that \nsga and \nsgad solutions are clustered closely, in contrast to the solutions obtained by \ras, which are more dispersed and offer better coverage of the DOI approximation in Plot~(a).
\\
\begin{table}[t]
\captionsetup{width=\textwidth}
\centering
\caption{Number of failures identified in the reference sets for the MNIST and AVP 
case studies across different test oracle definitions. The percentages of failures in the reference sets for AVP are based on 15,625 generated samples, and for MNIST, these percentages are based on 1,000 samples.}
\begin{tabular}{clccc}
\label{tab:num-failures}
 \textbf{Case Study}  & \textbf{Measure} & \textbf{\clarge} & \textbf{\cmid} & \textbf{\csmall} \\ \hline \\[-1em]
  AVP                 & \# Failures      & 558                   & 90   & 62 \\
                      & Percentage       & 3.6\%               & 0.6\%   & 0.4\%     \\ \hline\\[-1em]
  MNIST               & \# Failures      & 266               & 242              & 181         \\
                      & Percentage       & 26.6\%                & 24.2\%             & 18.1\%       \\ \bottomrule
\end{tabular}
\end{table}
We have also provided in Table~\ref{tab:num-failures} the number of failing tests in the reference set for \cmid for AVP, as well as the reference set sizes for MNIST. It can be observed that the number of failures identified  for the most constrained oracle definition is relatively small compared to the total number of samples (below 1\% for AVP and below 20\% for MNIST), indicating that the testing problems are not trivial.
 

\textbf{Clarification Regarding the Applicability of CID.} \emph{The \metric{} metric is not intended as a universal test coverage adequacy criterion applicable to various systems in practice. Instead, our metric acts as a quality indicator, assisting in the empirical assessment of testing approaches when an approximated reference set for the system under test is computable.} Our proposed metric, \metric{}, is explicitly designed to evaluate testing approaches during the development of an algorithm and is not meant for evaluating solutions in the application phase of the algorithm. 


Further, the validity of \metric{} values is tied to the accuracy of the reference set approximating the DOI.  Demonstrating the robustness of reference sets may become challenging for large, multi-dimensional search spaces (e.g., more than three dimensions), where one system execution may take minutes to hours. To mitigate this, we can use surrogate models~\citep{Raja16NeuralNSGA2,MatinnejadNBB14,NejatiSSFMM23,MatinnejadNBBP13} as approximations of the SUT, avoiding costly executions. Alternatively, we can evaluate a candidate search algorithm using \metric{}, calculated for mathematical test problems in SBST benchmarks~\citep{testfunctions}. Although these mathematical problems may not represent real-world ADS, they allow us to assess and potentially refute an ineffective candidate search algorithm. By augmenting these mathematical problems with a failure concept, i.e., a test oracle function, we can test the capabilities of a candidate search algorithm with respect to DOI coverage.

\textbf{Lesson Learned:} \textit{Pareto-based approaches perform better than na\"ive random testing in efficiently identifying single failures, yet they do not surpass random search in covering the areas of the search space that reveal failures.}  The results from RQ1 and RQ2 show that \nsga and \mopso cannot outperform randomized testing in terms of the DOI coverage. However, based on the results we obtained, we can confirm that \nsga and \mopso are on average better than \ras if the purpose is to find individual failures fast.

In particular, \autoref{tab:efficiency} shows the results of \nsga and \ras compared with respect to the first iteration where they can find a failing test input for the AVP case study. As shown in the table, on average, \nsga is able to find the first failure in less than 5 iterations for all test oracles, while \ras needs on average five or more iterations to find the first failure. 
The first iteration in which a search algorithm identifies a failure has been commonly used in literature to evaluate the efficiency of testing algorithms~\citep{Klück19Nsga2ADAS}.  
Our results confirm the superiority of \nsga over \ras when we consider the efficiency of finding failures. We note that this result does not contradict our results regarding the weakness of \nsga\ in covering failures, as covering the space of failure-inducing regions is a different objective from identifying several failure instances.\\
In addition, the results in \autoref{tab:efficiency} show that the AVP case study is a sufficiently difficult case study for meta-heuristic search algorithms, as \nsga already finds more failures than RS and does so faster. Demonstrating that \nsga outperforms RS in its optimization goal, which is finding failures in this case, is considered a sanity check comparison, indicating that the underlying case study system presents a sufficiently difficult problem warranting the use of a meta-heuristic algorithm.


\begin{table}[t]
\centering
    \captionsetup{width=1\textwidth}
\caption{AVP Case Study. The average iteration number in which the first failure-revealing test is identified by different algorithms, based on 10 runs. Pareto-based search algorithms outperform randomized search for the majority of all oracle types in finding the first failure.}\label{tab:efficiency}
\begin{tabular}{ c c c c}
\toprule
& \clarge & \cmid & \csmall\\
\hline
NSGA-II & 1.5 & 4.5 & 3.4 \\
NSGA-II-D  & 1.1 & 5.5 & 6.3  \\
\mopso & 1.8 & 7.4 & 7.9 \\
RS &  5 & 7 & 10 \\

\bottomrule
\end{tabular}
\end{table}
\textbf{Root Cause Analysis.} \textit{The motivation for our study is that the identification of diverse failing test inputs can support the localization of the root cause behind the identified failures. However, it is out of the scope of our study to analyze the underlying root causes of these failures.}
As demonstrated by the works from Jodat et al.~\citep{JodatSpurious23} and Abdessalam et al.~\citep{Raja18NSGA2DT} decision trees or decision rules can be used to derive failure explanations from a set of failure-revealing test inputs. For example, the work by Abdessalam et al. shows that failure-inducing test inputs can indicate that the reason for failure is due to environmental factors, such as high road curvature or hardware limitations, meaning an additional camera or one with a greater field of view might be required to capture vulnerable road users in sharp curves. Regardless, identifying failures is a prerequisite for fault localization and repair. Therefore, before proceeding with the fault localization and repair steps, we need testing techniques that can effectively identify as many diverse test inputs as possible, located in different areas of the search space that lead to failures.

\textbf{Feasibility and Accuracy of Reference Sets:} \textit{Regarding the computation of reference sets, we note that the higher the sampling rate, the higher the accuracy of the reference sets, but the more expensive it is to create those sets. We use three considerations to achieve a balance between the accuracy and feasibility of the reference sets' computations: (i) converging CID values, (ii) total time budget for reference set computation, and (iii) the sensitivity of the SUT's behavior to the sampling resolutions.}\\
i) We consider different sampling rates to see whether CID values stabilize and converge. Once we observe that the CID values stabilize at a specific sampling rate, we do not need to consider higher sampling rates since they are unlikely to significantly change the CID values. For instance, as shown in Table~\ref{tab:comparison_refset}, the CID values stabilize for all oracle definitions at a sampling rate of 25.\\
ii) We ensure that we have sufficient time to generate the reference sets. For instance, a sampling rate of 25 leads to the generation of 15,625 samples in the reference set of AVP with a three-dimensional test input space, which takes 86.8 hours. The number of samples increases exponentially with the number of input space dimensions. Hence, we consider capping the sampling rate based on our available time budget.\\
iii) We consider the sensitivity of the system's behavior to the selected sampling resolution. For instance, for the pedestrian velocity, which is one of AVP's input variables, a change of less than 0.075 m/s is unlikely to impact the system's behavior. Our sampling rate of 25 already samples this variable at a finer-grained resolution than 0.075 m/s. Therefore, we consider this sampling rate sufficient for computing stable CID values, making a higher resolution unnecessary.

\section{Conclusion and Future Work}
\label{sec:conclusion}
In this paper, we studied the ability of Pareto-based Search-Based Software Testing (SBST) to cover failure-revealing test inputs. Based on a theoretical argumentation and an empirical evaluation we have shown that Pareto-driven testing cannot achieve a high coverage of failures.

We introduced a new metric, \metric{}, to quantitatively evaluate the coverage of failure-inducing areas in a search domain and discussed its properties. 
In our empirical analysis using two case studies -- an industrial automated valet parking system and a handwritten digit classification system -- we demonstrate that the Pareto-based testing techniques \nsga and \mopso are less effective than random search in covering failure-revealing tests in 11 out of 12 alternative failure definitions across both studies, and in one case, all methods exhibited similar performance.

In addition, we demonstrated that augmenting \nsga with a diversified fitness function and a repopulation operator, adapted from a state-of-the-art testing approach does improve its performance in covering failure-revealing tests in four out of six alternative failure oracle definitions. However, in all comparisons, it does not perform better than random testing in terms of failure coverage.

Our investigation highlights the limitations of Pareto-driven testing in covering failure-revealing tests, emphasizing the need for practitioners using SBST with Pareto-based testing approaches to be aware of these constraints. Our observation could further confirm that Pareto-based testing is good in identifying failures fast, while for the coverage of failures, other innovative search strategies are necessary. 
Our future work is to combine randomized testing with machine learning models to improve the failure space coverage. The application of the \metric{} metric also confirms the importance of surrogate models and benchmarking systems to evaluate SBST algorithms, thereby avoiding expensive system executions.



\newpage
\appendix
\section{Proofs}

We provide the proofs for the following reformulated statements from Section~\ref{sec:metrics_qi}:

\begin{enumerate}
\item  Given an optimal reference set, the error of \metric{} reduces linearly with the maximal distance of adjacent points in the reference set decreasing.
    \item For an optimal reference set obtained with a uniform sampling approach, the CID's error tends to zero with the increasing number of points sampled.
\end{enumerate}

\subsection{Proof Statement 1}


The proofs are provided for $q = 1$ from the Definition of \metric{}. The proof for $q > 1$ can be obtained with similar derivations. We start with defining two reference sets $G$ and $F$, obtained by performing sampling within the domain $D\subseteq \mathbb{R}^n$ on a coarse and fine grids respectively, which are not necessarily structured. Let be $S$ the number of all separated regions of the DOI, then, since the reference sets are optimal, both reference sets witness every separated continuous region $r_s,\ s = 1, \ldots, S$. From the Definition of \metric{}, the following holds:
\begin{align}
    CID(A,Z) = \frac{1}{|Z|} \sum_{z \in Z} d_z = \frac{1}{|Z|} \sum_{s=1}^{S} \sum_{z \in  ( Z \cap r_s )} d_z
    \label{cid_def}
\end{align}
, i.e.,
\begin{align}
       CID(A,Z) =  \sum_{s=1}^{S} CID(A, Z \cap r_s)
       \label{cid_additivity}
\end{align}

Computing \metric{} for the test set A, using the reference sets $G$ and $F$, we obtain:
\begin{align}
    CID_G = \frac{1}{|G|} \sum_{i = 1}^{|G|} d_i^G 
    \label{cid_g_def}
\end{align}
\begin{align}
    CID_F = \frac{1}{|F|} \sum_{j = 1}^{|F|} d_i^F
    \label{cid_f_def}
\end{align}

From Equation~\ref{cid_additivity}, w.l.o.g., let's assume that $S = 1$, and let us denote $r_1$ as $r$.
The difference between the \metric{} values for the corresponding reference sets is then:

\begin{align}
    CID_F - CID_G = \frac{1}{|G|} \sum_{i = 1}^{|G|} d_i^G  - \frac{1}{|F|} \sum_{j = 1}^{|F|} d_i^F
    \label{cid_difference}
\end{align}

We partition the connected region $r$ using a Voronoi diagram built on $G$ into $|G|$ Voronoi cells $c_i$, $i = 1, \ldots, |G|$. Then, we assign to each reference point $f_j \in F$ a Voronoi cell $c_i$ if the reference point lies within the cell, and denote the reference point as $f_k^i$. If $f_j$ lies on a boundary of a cell, we assign it to any of the neighbouring cells. Let be $R^*$ the maximal distance of adjacent points in $G$. Each of the cells lies within a ball $B(g_i; R_i)$, and the radius of each of the balls is bounded by $R^*$, i.e. $ R_i \leq R^*$. As $f_k^i$ belongs to the cell $c_i$ with the center $g_i$, the distance between them is bounded: $\lVert \va*{\xi_{ik}} \rVert \leq R^*$,
where $\va*{\xi_{ik}}$ is the relative position between $f_k^i$, and $g_i$.

By applying the triangle equality of vector addition for the relative positions from the reference points $f_j \equiv f_i^k$ and $g_i$ to the test point yields:
\begin{align}
    \va*{d_j^F} =  \va*{d_i^G} + \va*{\xi_{ik}}
    \label{triangle_eq}
\end{align}
Using Equation ~\ref{cid_difference}, $CID_F$ from Equation ~\ref{cid_f_def} can be rewritten as:

\begin{align}
    \metric{}_F = \frac{1}{|F|} \sum_{i = 1}^{|G|} \sum_{k = 1}^{K_i} \lVert \va*{d_i^G} + \va*{\xi_{ik}}\rVert,
    \label{cid_f_parsed}
\end{align}
where $K_i$ is the number of points $f_k^i$ belonging to $c_i$.

With Equations~\ref{cid_f_parsed} and \ref{cid_difference}, the following holds:

\begin{align}
    CID_F - CID_G = \frac{1}{|F|} \sum_{i = 1}^{|G|} \sum_{k = 1}^{K_i} \lVert \va*{d_i^G} + \va*{\xi_{ik}}\rVert - \frac{1}{|G|} \sum_{i = 1}^{|G|} \lVert \va*{d_i^G}\rVert
    \label{cid_difference_parsed}
\end{align}

The norm $\lVert \cdot \rVert$ satisfies the following inequalities:

\begin{itemize}
    \label{ineq:triangle}
    \item Triangle inequality: $\lVert x + y \rVert \leq \lVert x \rVert + \lVert y \rVert$
    \label{ineq:rev-triangle}
    \item Reverse triangle inequality: $\lVert x + y \rVert \geq \lvert \lVert x\rVert - \lVert y \rVert \rvert$
\end{itemize}

Applying the triangle inequality to Equation \ref{cid_difference_parsed} yields:

\begin{equation}
\begin{aligned}
    \metric{}_F - \metric{}_G \leq 
    \frac{1}{|F|} \sum_{i = 1}^{|G|} \sum_{k = 1}^{K_i}\lVert \va*{d_i^G} \rVert - \frac{1}{|G|} \sum_{i = 1}^{|G|} \lVert \va*{d_i^G}\rVert
    + 
    \frac{1}{|F|} \sum_{i = 1}^{|G|}\sum_{k = 1}^{K_i} \lVert \va*{\xi_{ik}}\rVert \underset{*}{=} \\
    \frac{1}{|F|} \sum_{i = 1}^{|G|} (K_i - K^*) \lVert \va*{d_i^G}\rVert + \frac{1}{|F|} \sum_{i = 1}^{|G|}\sum_{k = 1}^{K_i} \lVert \va*{\xi_{ik}}\rVert 
    \underset{**}{\overset{|F| \rightarrow \infty}{\leq}} C(|G|) + R^*.
    \label{cid_difference_upper}
\end{aligned}
\end{equation}

Note:
\begin{itemize}
    \item [*]: For the equality, $K^*$ is the average number of points $f_k^i$ within a cell $c_i$, for $i = 1, \ldots |G|$, i.e. $K^* = {|F|}/{|G|}$. 
    \item [**]: For the inequality, presuming we have a uniform reference set $F$ of an infinite size, the first sum represents the deviation from the average of the volume of each Voronoi cell. As shown in \citep{GilbertCrystals62}, the deviation reduces with the increasing number of points. Furthermore, the reduction is even higher for higher dimensions \citep{DevroyeVoroinoCellsMeasure2015}. Note, that for a reference set obtained using a structured grid, the variance of the volume is equal to zero, as the volumes are equal for each cell, which means that the first sum is equal to zero. In other words, the first sum represents the uniform properties of the reference sets, and the more uniform the reference set, the smaller the variance is. We presume that the set of sampled points, and therefore the reference set $G$ are uniform enough to neglect the first sum. The second sum is bounded, as the size of each cell is bounded by the respective radius of a ball. 
\end{itemize}

Applying the reverse triangle inequality on Equation~\ref{cid_difference_parsed}, the following holds:

\begin{equation}
\begin{aligned}
      \metric{}_F - \metric{}_G \geq \frac{1}{|F|} \sum_{i = 1}^{|G|} \sum_{k = 1}^{K_i} | d_i^G - \lVert \va*{\xi_{ik}} \rVert| - \frac{1}{|G|}\sum_{i = 1}^{|G|} d_i^G \underset{*}{=} \\
 \frac{1}{|F|} \sum_{i = 1}^{|G|} \sum_{k = 1}^{K_i^+} (d_i^G - \lVert \va*{\xi_{ik}} \rVert)  + 
 \frac{1}{|F|} \sum_{i = 1}^{|G|} \sum_{k = 1}^{K_i^-} (\lVert \va*{\xi_{ik}} \rVert - d_i^G)  -
 \frac{1}{|F|} \sum_{i = 1}^{|G|} K^* d_i^G = \\
 \frac{1}{|F|} \sum_{i = 1}^{|G|} (K_i^+ - K^*) d_i^G  - \frac{1}{|F|} \sum_{i = 1}^{|G|} \sum_{k}^{K_i^+} \lVert \va*{\xi_{ik}} \rVert + 
 \frac{1}{|F|} \sum_{i = 1}^{|G|} \sum_{k}^{K_i^-} (\lVert \va*{\xi_{ik}} \rVert - d_i^G)
 \label{cid_difference_lower_p1}
\end{aligned}
\end{equation}

Note:
\begin{itemize}
    \item [*]: For the equality, we separate the sum of the moduli over the reference points within the cell into two sums: where the expression under the modulus is positive ($K_i^+$ terms), and negative ($K_i^-$ terms).
\end{itemize}

Now, we consider each of the sums from Equation~\ref{cid_difference_lower_p1} separately.
For the second sum holds:

\begin{equation}
\begin{aligned}
     \frac{1}{|F|} \sum_{i = 1}^{|G|} \sum_{k}^{K_i^+} \lVert \va*{\xi_{ik}} \rVert \leq 
     \frac{1}{|F|} \sum_{i = 1}^{|G|} \sum_{k}^{K_i} \lVert \va*{\xi_{ik}} \rVert  \leq
     R^*,
\end{aligned}
\end{equation}
as the size of each cell is bounded by the respective radius of a ball.


To assess the rest of the sums, let us study the dependence of $K_i^-$ and $K_i^+$ on $R^*$. For a fixed test set, the following holds:
\begin{align}
    \exists G_0 : \forall G : |G| > |G_0| \implies K_i^- \leq R^*\cdot K^*.
    \label{k_minus}
\end{align}
As $K_i^+ + K_i^- = K_i$, and for a uniform structured grid, $K_i = K^*$, the following holds:
\begin{align}
K_i^+ = K^* - K_i^- \geq K^* - R^* \cdot K^* = (1 - R^*) \cdot K^*
\label{k_plus}
\end{align}

If a grid is not structured, the additional terms involving the variances of the cell volume can be neglected similarly to Equation ~\ref{cid_difference_upper}. Then, from Equation~\ref{k_minus}, for the third sum holds:

\begin{align}
     \frac{1}{|F|} \sum_{i = 1}^{|G|} \sum_{k}^{K_i^-} (\lVert \va*{\xi_{ik}} \rVert - d_i^G) \leq C' \cdot \frac{1}{K^*} \cdot R^* \cdot K^* =
     C^{'} \cdot R^* ;\ C'> 0.
\end{align}

Similarly, for the first sum in the Equation~\ref{k_plus} holds:

\begin{align}
    \left| \frac{1}{|F|} \sum_{i = 1}^{|G|} (K_i^+ - K^*) \cdot d_i^G \right| \leq C'' \cdot \frac{1}{K^*} \cdot R^* \cdot K^* = 
    C'' \cdot R^* ;\ C''> 0.
\end{align}

Then for the Equation~\ref{cid_difference_lower_p1} holds:
\begin{align}
    CID_F - CID_G \geq - (C' + C'' + 1) \cdot R^* \geq - C''' \cdot R^*;\ C'''> 0.
    \label{cid_difference_lower_p2}
\end{align}

Finally, from Equation~\ref{cid_difference_lower_p2} and Equation~\ref{cid_difference_upper} follows:

\begin{align}
    | CID_F - CID_G | \leq C^* \cdot R^*,  \  C^* = \max(C''', C).
    \label{cid_linear}
\end{align}

Equation~\ref{cid_linear} implies that the error reduces linearly with the maximal distance between adjacent points in the reference set decreasing, which proofs the Statement 1.


\subsection{Proof Statement 2}

Because the sampling strategy is uniform, the maximum distance between adjacent points in the reference set $R^*$ decreases with the increasing number of sampled points. Statement 2 follows from Equation~\ref{cid_linear}. 




 
\section*{Acknowledgments}
\includegraphics[scale=0.018]{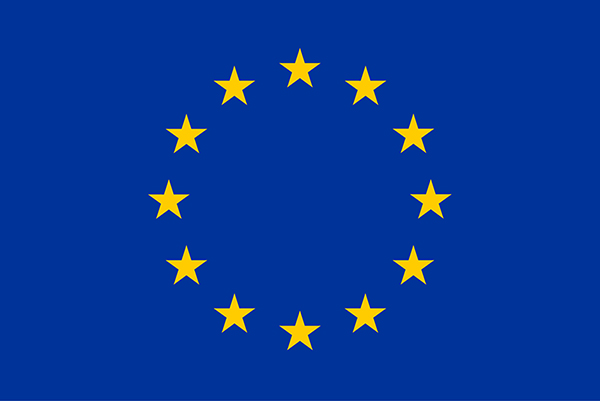}  This paper has received funding from the European Union’s Horizon 2020 research and innovation programme under grant agreement No 956123, and the NSERC of Canada under the discovery grant.  We sincerely appreciate the valuable comments and suggestions from the reviewers and colleagues which supported our research.

\section*{Data availability}
\label{sec:data}
Our online material~\cite{replication-package} includes the implementation of our case studies, ~scripts and description for performing a coverage analysis on the MNIST case study, and experimental results, diagrams and statistical test results for the experiments on both case studies. Material for replicating the results for the AVP case study could not have been disclosed, as the system under test of this study is a proprietary system of the industrial partner.

\section*{Declarations}

\subsection*{Conflicts of interests}
\label{sec:coi}
The authors have no competing interests to declare that are relevant to the content of this article.

\balance
\bibliography{papers}

\end{document}